\documentclass[pra,preprint,superscriptaddress,floatfix,showkeys,showpacs]{revtex4-1}

\usepackage{amsmath,amsfonts,amssymb}
\usepackage{graphicx,color}

\def\beq{\begin{equation}}
\def\eeq{\end{equation}}
\def\beqn{\begin{eqnarray}}
\def\eeqn{\end{eqnarray}}

\def\r {{\bf r}}

\def\Q {{\bf Q}}

\def\r {{\bf r}}

\def\Q {{\bf Q}}

\begin{document}

\title{Analysis of a trapped Bose-Einstein condensate in terms of position, momentum, and angular-momentum variance}
\author{Ofir E. Alon}
\email{ofir@research.haifa.ac.il}
\affiliation{Department of Mathematics, University of Haifa, Haifa 3498838, Israel}
\affiliation{Haifa Research Center for Theoretical Physics and Astrophysics, University of Haifa,
Haifa 3498838, Israel}

\begin{abstract}
We analyze, analytically and numerically, 
the position, momentum, and in particular the angular-momentum variance 
of a Bose-Einstein condensate (BEC) trapped in a two-dimensional anisotropic trap
for static and dynamic scenarios.
The differences between the variances at the mean-field level,
which are attributed to the shape of the BEC,
and the variances at the many-body level,
which incorporate depletion, 
are used to characterize position, momentum, and angular-momentum correlations in the BEC
for finite systems and at the limit of an infinite number of particles
where the bosons are $100\%$ condensed.
Finally, we also explore inter-connections between
the variances.
\end{abstract}


\maketitle 

\section{Introduction}\label{INTRO}

Bose-Einstein condensates (BECs) made of ultra-cold atoms 
offer a wide platform to study many-body physics \cite{rev1,rev2,rev3,rev4,rev5}.
Here, there is a growing interest in the so-called infinite-particle limit \cite{INF1,INF2,INF3,INF4,INF5,INF6,INF7,INF8,INF9,INF10,INF11},
in which the interaction parameter (i.e., the product of the interaction strength times the number of particles)
is kept fixed while the number of particles is increased to infinity.
At the infinite-particle limit, the energy per particle, density per particle, and reduced density matrices \cite{RDM_book} per particle
computed at the many-body level of theory boil down to those obtained in mean-field theory \cite{INF2,INF3,INF4,INF5,INF9,INF11},
despite the fact that the respective many-boson wavefunctions are (much) different \cite{INF8,INF10}.
It turns out that variances of many-particle operators are a useful tool
to characterize correlations (namely, differences between respective many-body and mean-field quantities)
that exist even when the interacting bosons are 100\% condensed \cite{INF6,INF7}.

The variance of a many-particle operator of a trapped BEC generally depends
on the trap shape, strength and sign of the interaction and,
in out-of-equilibrium problems, on time.
Consequently, the difference between variances computed at the many-body and mean-field levels of theory
also depends on these variables and, of course, on the observable under examination.
The first examples \cite{INF6,INF7} concentrated on one-dimensional problems
and the position and momentum variances,
and investigated conditions and mechanisms for the differences between
the respective many-body and mean-field variances at the infinite-particle limit.
In two spatial dimensions, further types of trap topologies come into play,
and respective many-body and mean-field variances can exhibit additional phenomena,
such as opposite anisotropy \cite{var1} and distinct (effective) dimensionality \cite{var2}. 
The many-body variance of a trap BEC
has been applied to extract excitations \cite{var_ap1},
analyze the range of inter-particle interaction \cite{var_ap2},
examine the effects of asymmetry of a double-well potential \cite{var_ap3},
and to assess numerical convergence \cite{conv_1,conv_2}.

So far, only the position and momentum variances were studied for BECs in rather general traps.
In \cite{var_am1,var_am2},
the angular-momentum variance is studied for BECs in two-dimensional isotropic traps,
and scenarios were the mean-field angular-momentum variance has less \cite{var_am1} or more \cite{var_am2}
symmetry (in term of its conservation) than the many-body angular-momentum variance are identified.
Going beyond these works,
in the present work we study,
analytically and numerically, 
the angular-momentum variance 
of a trapped BEC in a two-dimensional anisotropic trap
for static and dynamic scenarios,
and analyze the difference between the many-body and mean-field variances
for finite systems and at the limit of an infinite number of particles.
Furthermore,
we also study the respective position and momentum variances,
and thereby offer a comprehensive characterization of the BEC in terms of its variances.
This would allow us to put forward inter-connections between the variances.

The structure of the paper is as follows.
In Sec.~\ref{ANI_HIM} we study the position, momentum, and angular-momentum
variances of the ground state within an exactly-solvable model, the anisotropic harmonic-interaction model.
In Sec.~\ref{ANN_TILT} we study numerically the time-dependent variances of
an out-of-equilibrium BEC sloshing in a tilted annulus.
Summary and outlook are given in Sec.~\ref{SUMM}.
Finally, Appendix \ref{VAR_TRANS} discusses
translations of variances and inter-connections of the latter.

\section{The anisotropic harmonic-interaction model}\label{ANI_HIM}

Solvable models of particles interacting by harmonic forces, or, briefly, the harmonic-interaction model (and its variants),
have drawn much attention \cite{HIM1,HIM2,HIM3,HIM4,HIM5,HIM6,HIM7,HIM8,HIM9,HIM10,HIM11,HIM12,HIM13,HIM14,HIM15}.
Here we consider the anisotropic two-dimensional harmonic-interaction model
\beqn\label{Ham_HIM_Ani}
& & \hat H(\r_1,\ldots,\r_N) = \sum_{j=1}^N \left[\left( -\frac{1}{2}\frac{\partial^2}{\partial x_j^2} + \frac{1}{2}\omega_x^2 x_j^2\right) + 
\left(-\frac{1}{2}\frac{\partial^2}{\partial y_j^2} + \frac{1}{2}\omega_y^2 y_j^2 \right)\right] + \nonumber \\
& & \qquad + \lambda_0 \sum_{1 \le j < k}^N \left[(x_j-x_k)^2+(y_j-y_k)^2\right], \
\eeqn
where $\lambda_0$ is the interaction strength; positive values imply attraction and negative repulsion.
Without loss of generality we take $\omega_y>\omega_x$, namely,
that the trap is tighter along the y-axis than along the x-axis
(the trap anisotropy satisfies $\frac{\omega_y}{\omega_x}>1$).
Here and hereafter $\hbar=m=1$.

Transforming from Cartesian to Jacobi coordinates,
\beqn\label{Jacobi}
& & Q_{k,x} = \frac{1}{\sqrt{k(k+1)}} \sum_{j=1}^k \left(x_{k+1}-x_j\right), \quad
Q_{k,y} = \frac{1}{\sqrt{k(k+1)}} \sum_{j=1}^k \left(y_{k+1}-y_j\right), \quad
1 \le k \le N-1, \nonumber \\
& & Q_{N,x} = \frac{1}{\sqrt{N}} \sum_{j=1}^N x_j, \quad
Q_{N,y} = \frac{1}{\sqrt{N}} \sum_{j=1}^N y_j, \
\eeqn
the many-body solution for the ground state is given by
\beqn\label{MB_PSI}
& & \Psi(\Q_1,\ldots,\Q_N) = \left(\frac{\omega_x}{\pi}\right)^{\frac{1}{4}} \left(\frac{\omega_y}{\pi}\right)^{\frac{1}{4}}
\left(\frac{\Omega_x}{\pi}\right)^{\frac{N-1}{4}} \left(\frac{\Omega_y}{\pi}\right)^{\frac{N-1}{4}} \times \nonumber \\
& & \qquad \times e^{-\frac{1}{2} \left( \Omega_x \sum_{j=1}^{N-1} Q_{j,x}^2 + \omega_x Q_{N,x}^2 \right)} \times
e^{-\frac{1}{2} \left( \Omega_y \sum_{j=1}^{N-1} Q_{j,y}^2 + \omega_y Q_{N,y}^2 \right)} = \nonumber \\
& & = \Psi(\r_1,\ldots,\r_N) = \left(\frac{\omega_x}{\pi}\right)^{\frac{1}{4}} \left(\frac{\omega_y}{\pi}\right)^{\frac{1}{4}}
\left(\frac{\Omega_x}{\pi}\right)^{\frac{N-1}{4}} \left(\frac{\Omega_y}{\pi}\right)^{\frac{N-1}{4}} \times \nonumber \\
& & \qquad \times e^{-\frac{\alpha_x}{2} \sum_{j=1}^N x_j^2 - \beta_x \sum_{1 \le j < k}^N x_j x_k} \times
e^{-\frac{\alpha_y}{2} \sum_{j=1}^N y_j^2 - \beta_y \sum_{1 \le j < k}^N y_j y_k}, \
\eeqn
where
\beq\label{Rel_Omega}
\Omega_x=\sqrt{\omega_x^2+2N\lambda_0}, \qquad
\Omega_y=\sqrt{\omega_y^2+2N\lambda_0}
\eeq
are the interaction-dressed frequencies of the relative-motion degrees-of-freedom,
and
\beqn\label{Alpha_Beta_PAR}
& & \alpha_x = \Omega_x+\beta_x,
\qquad \beta_x=\frac{1}{N}\left(\omega_x-\Omega_x\right), \nonumber \\
& & \alpha_y= \Omega_y+\beta_y,
\qquad \beta_y=\frac{1}{N}\left(\omega_y-\Omega_y\right), \
\eeqn
are parameters arising in the transformation back from Jacoby to Cartesian coordinates.
Eq.~(\ref{Rel_Omega}) prescribes the range of interactions for which the system is trapped,
$\lambda_0 > - \frac{\omega_x^2}{2N}$,
i.e., from moderate repulsion to any attraction.
Clearly, the many-body solution (\ref{MB_PSI}) in two spatial dimensions 
factorizes to a product of respective one-dimensional many-body solutions. 

All properties of the ground state can in principle be obtained from $\Psi$, such as the energy, densities, and reduced density matrices,
see \cite{HIM3}.
Here, as mentioned above, we concentrate on variances and their inter-connections. 
The many-particle position $\hat X=\sum_{j=1}^N x_j$, $\hat Y=\sum_{j=1}^N y_j$ variance per particle is given by 
\beq\label{Var_X_Y}
 \frac{1}{N}\Delta^2_{\hat X} = \frac{1}{2\omega_x}, \qquad \frac{1}{N}\Delta^2_{\hat Y} = \frac{1}{2\omega_y}.
\eeq
Due to the symmetry of center-of-mass separation in the Hamiltonian (\ref{Ham_HIM_Ani}),
the many-particle position variance per particle is independent both of the interaction strength and 
the number of
bosons in the system.
Similarly,
the many-particle momentum $\hat P_X=\sum_{j=1}^N \frac{1}{i} \frac{\partial}{\partial x_j}$, 
$\hat P_Y=\sum_{j=1}^N \frac{1}{i} \frac{\partial}{\partial y_j}$ variance per particle is given by 
\beq\label{Var_PX_PY}
 \frac{1}{N}\Delta^2_{\hat P_X} = \frac{\omega_x}{2},
\qquad \frac{1}{N}\Delta^2_{\hat P_Y} = \frac{\omega_y}{2},
\eeq
reflecting the minimal uncertainty product
$\frac{1}{N}\Delta^2_{\hat X}\frac{1}{N}\Delta^2_{\hat P_X}=\frac{1}{N}\Delta^2_{\hat Y}\frac{1}{N}\Delta^2_{\hat P_Y}=\frac{1}{4}$
of the interacting system in the anisotropic harmonic trap.

The many-particle angular-momentum 
$\hat L_Z=\sum_{j=1}^N \frac{1}{i}\left( x_j \frac{\partial}{\partial y_j} - y_j \frac{\partial}{\partial x_j}\right)$ variance per particle
is, at least for bosons, a less familiar and more intricate quantity.
After some lengthy
but otherwise straightforward algebra it is given by
\beqn\label{Var_Lz}
& & \frac{1}{N}\Delta^2_{\hat L_Z} = \frac{1}{4} \frac{\left(\Omega_y-\Omega_x\right)^2}{\Omega_y\Omega_x}
\left(\frac{N-1}{N}\right)^2 \Bigg[\left(1+\frac{1}{N-1}\frac{\Omega_y}{\omega_y}\right)
\left(1+\frac{1}{N-1}\frac{\Omega_x}{\omega_x}\right) + \\
& & \ \ + \left(\frac{\Omega_y}{\omega_y}-1\right)\left(\frac{\Omega_x}{\omega_x}-1\right)\Bigg] + 
\frac{1}{4N}\frac{\left[\left(\omega_y-\Omega_y\right)-\left(\omega_x-\Omega_x\right)\right]
\left[\left(\omega_y+\Omega_y\right)-\left(\omega_x+\Omega_x\right)\right]}{\omega_y\omega_x}, \ \nonumber
\eeqn
where we have made use of the bosonic permutational symmetry, the structure of $\Psi$,
\beq\label{Lz_Psi}
\hat L_Z \Psi = - \frac{1}{i} \left\{\left[(\alpha_y-\beta_y)-(\alpha_x-\beta_x)\right] \left(\sum_{j=1}^N x_jy_j\right) + 
(\beta_y-\beta_x) \left(\sum_{j=1}^N x_j\right) \left(\sum_{k=1}^N y_k\right) \right\} \Psi,
\eeq
and the inverse coordinate transformations
\beqn\label{Jac_Car}
& & x_N=\frac{1}{\sqrt{N}} Q_{N,x}+\sqrt{\frac{N-1}{N}}Q_{N-1,x}, \qquad
y_N=\frac{1}{\sqrt{N}} Q_{N,y}+\sqrt{\frac{N-1}{N}}Q_{N-1,y}, \nonumber \\
& & x_{N-1}=\frac{1}{\sqrt{N}}Q_{N,x}-\frac{1}{\sqrt{N(N-1)}}Q_{N-1,x}+\sqrt{\frac{N-2}{N-1}}Q_{N-2,x}, \nonumber \\
& & y_{N-1}=\frac{1}{\sqrt{N}}Q_{N,y}-\frac{1}{\sqrt{N(N-1)}}Q_{N-1,y}+\sqrt{\frac{N-2}{N-1}}Q_{N-2,y} \
\eeqn
to evaluate the various integral terms contributing to (\ref{Var_Lz}).

The angular-momentum variance per particle of the ground state (\ref{MB_PSI})
depends on the dressed frequencies, $\Omega_x$ and $\Omega_y$, and the number of particles $N$.
Namely, unlike the respective position and momentum variances
it depends explicitly on the interaction strength and the number of particles.
$\frac{1}{N}\Delta^2_{\hat L_Z}$ is, of course, non-zero only for anisotropic traps
[for isotropic traps, from $\omega_y=\omega_x$ we get $\Omega_y=\Omega_x$ and expression (\ref{Var_Lz}) then vanishes].
For non-interacting bosons,
Eq.~(\ref{Var_Lz}) boils down to
$\frac{1}{N}\Delta^2_{\hat L_Z}=\frac{1}{4}\frac{(\omega_y-\omega_x)^2}{\omega_y\omega_x}=
\frac{1}{4}\frac{\left(\frac{\omega_y}{\omega_x}-1\right)^2}{\frac{\omega_y}{\omega_x}}$,
the value for a single particle in the anisotropic trap $\frac{1}{2}\omega_x^2 x^2 + \frac{1}{2}\omega_y^2 y^2$,
which only depends on the trap anisotropy.
Opposite to the non-vanishing of the angular-momentum variance,
we note that the expectation value of the angular momentum operator,
$\frac{1}{N}\langle\Psi|\hat L_Z|\Psi\rangle$,
vanishes for any anisotropy $\frac{\omega_y}{\omega_y}$, interaction strength $\lambda_0$, and number of particles $N$.
This is straightforward to see since $\Psi$ is even under reflection of all coordinates
$X \to -X$ and separately of $Y \to -Y$,
whereas $\hat L_Z$ is odd under reflection.

The anisotropic harmonic-interaction model (\ref{Ham_HIM_Ani}) can be solved analytically at the mean-field
level of theory as well, like in \cite{HIM3}, also see \cite{HIM15}.
Starting from the ansatz where each and every boson resides
in one and the same orbital,
the mean-field solution is given by
\beqn\label{GP_PSI}
& & \Phi^{GP}(\r_1,\ldots,\r_N) = \nonumber \\
& & = \left(\frac{\sqrt{\omega_x^2+2\Lambda}}{\pi}\right)^{\frac{N}{4}} 
\left(\frac{\sqrt{\omega_y^2+2\Lambda}}{\pi}\right)^{\frac{N}{4}}
 e^{-\frac{1}{2} \sqrt{\omega_x^2+2\Lambda}\sum_{j=1}^N x_j^2} \times
e^{-\frac{1}{2} \sqrt{\omega_y^2+2\Lambda}\sum_{j=1}^N y_j^2} = \nonumber \\
& & = \Phi^{GP}(\Q_1,\ldots,\Q_N) = \nonumber \\
& & = \left(\frac{\sqrt{\omega_x^2+2\Lambda}}{\pi}\right)^{\frac{N}{4}} 
\left(\frac{\sqrt{\omega_y^2+2\Lambda}}{\pi}\right)^{\frac{N}{4}}
e^{-\frac{1}{2} \sqrt{\omega_x^2+2\Lambda}\sum_{k=1}^N Q_{k,x}^2} \times
e^{-\frac{1}{2} \sqrt{\omega_y^2+2\Lambda}\sum_{k=1}^N Q_{k,y}^2}, \
\eeqn
where $\Lambda=(N-1)\lambda_0$ is the interaction parameter,
and $\Lambda>-\frac{\omega_x^2}{2}$ the condition for a trapped solution.
Like the many-body solution (\ref{MB_PSI}),
the mean-field solution (\ref{GP_PSI}) in two spatial dimensions 
factorizes to a product of respective one-dimensional mean-field solutions. 

The many-particle position variance computed at the mean-field level is given by 
\beq\label{Var_X_Y_GP}
 \frac{1}{N}\Delta^2_{\hat X,GP} = \frac{1}{2\sqrt{\omega_x^2+2\Lambda}},
\qquad \frac{1}{N}\Delta^2_{\hat Y,GP} = \frac{1}{2\sqrt{\omega_y^2+2\Lambda}},
\eeq
and seen to be dressed by the interaction.
Similarly, the many-particle momentum variance computed at the mean-field level is dressed by the interaction and given by 
\beq\label{Var_PX_PY_GP}
 \frac{1}{N}\Delta^2_{\hat P_X,GP} = \frac{\sqrt{\omega_x^2+2\Lambda}}{2},
\qquad \frac{1}{N}\Delta^2_{\hat P_Y,GP} = \frac{\sqrt{\omega_y^2+2\Lambda}}{2}.
\eeq
Interestingly, because the mean-field solution (\ref{GP_PSI}) is made of Gaussian functions,
it satisfies the minimal uncertainty product
$\frac{1}{N}\Delta^2_{\hat X, GP}\frac{1}{N}\Delta^2_{\hat P_X, GP}=
\frac{1}{N}\Delta^2_{\hat Y, GP}\frac{1}{N}\Delta^2_{\hat P_Y, GP}=\frac{1}{4}$
as well.
 
The many-particle angular-momentum variance computed at the mean-field level is given by
\beq\label{Var_Lz_GP}
 \frac{1}{N}\Delta^2_{\hat L_Z,GP} = \frac{1}{4} \frac{\left(\sqrt{\omega_y^2+2\Lambda}-\sqrt{\omega_x^2+2\Lambda}\right)^2}{\sqrt{\omega_y^2+2\Lambda} \sqrt{\omega_x^2+2\Lambda}},
\eeq
where we have made use of the structure and symmetries of $\Phi^{GP}$,
\beq
\hat L_Z \Phi^{GP} = - \frac{1}{i} \left(\sqrt{\omega_y^2+2\Lambda}-\sqrt{\omega_x^2+2\Lambda}\right) \sum_{j=1}^N x_j y_j \Phi^{GP},
\eeq
to arrive at the final expression.

The relation between the mean-field and many-body variances deserves a discussion.
Their difference is used to define position, momentum, and angular-momentum correlations in the system. 
For the position and momentum variances,
the following ratios hold,
\beqn\label{X_Y_PX_PY_Rat}
& & \frac{\frac{1}{N}\Delta^2_{\hat X,GP}}{\frac{1}{N}\Delta^2_{\hat X}}=\frac{1}{\sqrt{1+\frac{2\Lambda}{\omega_x^2}}}, \qquad
\frac{\frac{1}{N}\Delta^2_{\hat Y,GP}}{\frac{1}{N}\Delta^2_{\hat Y}}=\frac{1}{\sqrt{1+\frac{2\Lambda}{\omega_y^2}}}, \nonumber \\
& & \frac{\frac{1}{N}\Delta^2_{\hat P_X,GP}}{\frac{1}{N}\Delta^2_{\hat P_X}}=\sqrt{1+\frac{2\Lambda}{\omega_x^2}}, \qquad
\frac{\frac{1}{N}\Delta^2_{\hat P_Y,GP}}{\frac{1}{N}\Delta^2_{\hat P_Y}}=\sqrt{1+\frac{2\Lambda}{\omega_y^2}}, \
\eeqn
obviously
for any number of particles $N$.
These ratios simply imply that, since repulsion ($\Lambda<0$) broadens the position density,
the many-body position variance is smaller than the corresponding mean-field one for repulsive interaction, and vise verse for attraction ($\Lambda>0$).
Inversely, since repulsion narrows the momentum density,
the many-body momentum variance is larger than the corresponding mean-field one for repulsive interaction, and vise versa for attraction.
Furthermore,
both the position and momentum variances per particle exhibit the same anisotropies as the respective densities for any interaction parameter $\Lambda$,
namely,
if $\frac{1}{N}\Delta^2_{\hat X, GP} > \frac{1}{N}\Delta^2_{\hat Y, GP}$ then $\frac{1}{N}\Delta^2_{\hat X} > \frac{1}{N}\Delta^2_{\hat Y}$
is satisfied and, analogously,
if $\frac{1}{N}\Delta^2_{\hat P_X, GP} < \frac{1}{N}\Delta^2_{\hat P_Y, GP}$ then $\frac{1}{N}\Delta^2_{\hat P_X} < \frac{1}{N}\Delta^2_{\hat P_Y}$
is satisfied.
We shall return to these relations
and the anisotropy of the variance
in the numerical example below.

We now extend the above discussion to the infinite-particle limit,
in which the energy per particle, densities per particle, and reduced densities per particle 
at the mean-field and many-body levels of theory coincide, see in the context of the harmonic-interaction model \cite{INF11}.
Particularly, the system of bosons becomes $100\%$ condensed.
The results (\ref{X_Y_PX_PY_Rat}) for the position and momentum variances hold at the infinite-particle limit as well,
owing to the center-of-mass separability for any number of particles,
namely,
$\frac{\lim_{N \to \infty}\frac{1}{N}\Delta^2_{\hat X,GP}}{\lim_{N \to \infty}\frac{1}{N}\Delta^2_{\hat X}}=\frac{1}{\sqrt{1+\frac{2\Lambda}{\omega_x^2}}}$,
$\frac{\lim_{N \to \infty}\frac{1}{N}\Delta^2_{\hat Y,GP}}{\lim_{N \to \infty}\frac{1}{N}\Delta^2_{\hat Y}}=\frac{1}{\sqrt{1+\frac{2\Lambda}{\omega_y^2}}}$,
$\frac{\lim_{N \to \infty}\frac{1}{N}\Delta^2_{\hat P_X,GP}}{\lim_{N \to \infty}\frac{1}{N}\Delta^2_{\hat P_X}}=
\sqrt{1+\frac{2\Lambda}{\omega_x^2}}$, and
$\frac{\lim_{N \to \infty}\frac{1}{N}\Delta^2_{\hat P_Y,GP}}{\lim_{N \to \infty}\frac{1}{N}\Delta^2_{\hat P_Y}}=
\sqrt{1+\frac{2\Lambda}{\omega_y^2}}$.
For the angular-momentum variance the limit has to be taken explicitly
for each of the terms in (\ref{Var_Lz}). 
First are the frequencies (\ref{Rel_Omega}),
for which we have at the limit of an infinite number of bosons when $\Lambda$ is held fixed
\beq\label{INF_Omega}
\lim_{N \to \infty} \Omega_x = \sqrt{\omega_x^2+2\Lambda}, \qquad
\lim_{N \to \infty} \Omega_y = \sqrt{\omega_y^2+2\Lambda}.
\eeq
Then, the angular-momentum variance takes on the appealing form
\beq\label{Var_Lz_INF}
\lim_{N \to \infty} \frac{1}{N}\Delta^2_{\hat L_Z} = \frac{1}{4} 
\frac{\left(\sqrt{\omega_y^2+2\Lambda}-\sqrt{\omega_x^2+2\Lambda}\right)^2}{\sqrt{\omega_y^2+2\Lambda}\sqrt{\omega_x^2+2\Lambda}}
\left[1 + \left(\sqrt{1+\frac{2\Lambda}{\omega_y^2}}-1\right)\left(\sqrt{1+\frac{2\Lambda}{\omega_x^2}}-1\right)\right].
\eeq
Comparing (\ref{Var_Lz_INF}) to the mean-field expression (\ref{Var_Lz_GP}),
it is instrumental to prescribe their ratio
at the limit of an infinite number of particles
(where, as mentioned above, the density per particle and other properties coincide),
\beq\label{Lz_Rat_INF}
 \frac{\lim_{N \to \infty} \frac{1}{N}\Delta^2_{\hat L_{Z,GP}}}{\lim_{N \to \infty} \frac{1}{N}\Delta^2_{\hat L_Z}} =
\frac{1}{1 + \left(\sqrt{1+\frac{2\Lambda}{\omega_y^2}}-1\right)\left(\sqrt{1+\frac{2\Lambda}{\omega_x^2}}-1\right)},
\eeq 
which is always smaller than $1$ for interacting bosons in the anisotropic trap.
Furthermore, we see that for attractive interaction 
the many-body variance can become much larger than the mean-field quantity
in the anisotropic trap, signifying the growing necessity of the many-body treatment,
even when the system is $100\%$ condensed.
This concludes our investigation of a solvable anisotropic many-boson model in which
the variances of the momentum, position, and angular-momentum many-particle operators can
be computed and investigated analytically,
and their values at the many-body and mean-field levels of theory
compared and contrasted.

\section{Bosons in an annulus subject to a tilt}\label{ANN_TILT}

In most scenarios of interest, the position, momentum, and angular-momentum variance cannot be computed analytically.
This is in many cases the situations when symmetries are lifted.
Moreover, even when the variances can be compute for the ground state,
like in the previous section \ref{ANI_HIM},
their values for an out-of-equilibrium scenario are rarely within analytical reach.
This would be the situation of the present investigation.

Bosons in rings, annuli, and shells have attracted considerable attention \cite{rn1,rn2,rn3,rn4,rn5,rn6,rn7,rn8,rn9,rn10,rn11,rn12,rn13,rn14,rn15,rn16,rn17,rn18,rn19,rn20,rn21,rn22,rn23,rn24,rn25}.
Here we consider weakly-interacting bosons initially prepared in the ground state of a two-dimensional annulus.
The annulus is then suddenly slightly tilted,
leading to an out-of-equilibrium dynamics in an anisotropic setup.
We build on and extend the study of bosons' dynamics in an annulus within an isotropic setup \cite{var2}
(for which, e.g., the angular-momentum variance is $0$).
We analyze the BEC dynamics in terms of its time-dependent variances and other quantities of relevance,
see Figs.~\ref{f1}-\ref{f7} below.

We consider the out-of-equilibrium dynamics governed by
the many-particle Schr\"odinger equation in two spatial dimensions,
$\hat H(\r_1,\ldots,\r_N)\Psi(\r_1,\ldots,\r_N;t)=i\frac{\partial\Psi(\r_1,\ldots,\r_N;t)}{\partial t}$.
The bosons are initially prepared in the ground state of the annulus, see Fig.~1 in \cite{var2}.
The trap potential is given by $\hat V(\r)=0.05\r^4+V_0e^{-\frac{\r^2}{2}}$,
with barrier of heights $V_0=5$ and $10$ throughout this work.
The interaction between the bosons is repulsive and taken to be 
$\lambda_0W(\r-\r')=\lambda_0e^{-\frac{(\r-\r')^2}{2}}$, 
where the interaction strengths are $\lambda_0=0.02$ and $0.04$ throughout this work.
The form and extant of the interaction potential do not
have a qualitative influence on the physics to be described below.
At time $t=0$ a linear term is added such that
$V(\r)=0.05\r^4+V_0e^{-\frac{\r^2}{2}}+0.01 x$.
The physical meaning of the added potential is that a constant force pointing to the left is
suddenly acting on the interacting bosons.
Geometrically, the annulus can be considered to be slightly tilted to the left.
Symmetry-wise, the isotropy of the potential is lifted and anisotropy sets in.
All in all, the interacting bosons are not in their ground state any more,
and out-of-equilibrium dynamics emerges.

To compute the time-dependent many-boson wavefunction
we use the multiconfigurational time-dependent Hartree 
for bosons (MCTDHB) method \cite{MCTDHB1,MCTDHB2,MCTDHB3}.
MCTDHB represents the wavefunction
as a variationally-optimal ansatz 
which is a linear-combination
of all time-dependent permanents generated  
by distributing the $N$ bosons over $M$ time-adaptive orbitals.
The quality of the wavefunction increases with $M$, and convergence of quantities of interest attained.
The theory, applications, benchmarks, and extensions of MCTDHB are
extensively discussed in the literature, see, e.g., 
Refs.~\cite{BS1,BS2,BS3,BS4,BS5,BS6,BS7,BS8,BS9,BS10,BS11,BS12,BS13,BS14,BS15,BS16,BS17,BS18,BS19,BS20,BS21,BS22,BS23,BS24,BS25,BS26}.
Here we employ the numerical implementation in \cite{PACK1,PACK2} both for preparing the ground state \cite{MCHB}
(using imaginary-time propagation) and real-time dynamics.
Finally, we mention that MCTDHB is the bosonic version of the nearly-three-decades-established
distinguishable-particle multiconfigurational time-dependent Hartree method frequently used (alongside its extensions) in molecular physics 
\cite{MC1,MC2,MC3,MC4,MC5,MC6,MC7}.

From the time-dependent wavefunction $\Psi(\r_1,\ldots,\r_N;t)$, here normalized to $1$,
we compute properties of interest.
The reduced one-particle density matrix is defined as
$\rho(\r,\r';t)=N\int d\r_2 \cdots d\r_N\Psi^\ast(\r',\r_2,\ldots,\r_N;t)\Psi(\r,\r_2,\ldots,\r_N;t)=
\sum_j n_j(t) \phi^\ast_j(\r';t) \phi_j(\r;t)$,
where\break\hfill $\{\phi_j(\r;t)\}$ are the natural orbitals and $\{n_j(t)\}$ the natural occupations.
The number of particles residing outside the condensed mode $\phi_1(\r;t)$,
i.e., the total number of depleted particles,
is given by $\sum_{j>1} n_j(t) =N-n_1(t)$.
Analogously, the reduced two-particle density matrix is given by
$\rho(\r_1,\r_2,\r'_1,\r'_2;t)=N(N-1)\int d\r_3 \cdots d\r_N\Psi^\ast(\r'_1,\r'_2,\r_3,\ldots,\r_N;t)\times$\break\hfill
$\Psi(\r_1,\r_2,\r_3,\ldots,\r_N;t)=\sum_{jpkq} \rho_{jpkq}(t) \phi^\ast_j(\r'_1;t)\phi^\ast_p(\r'_2;t)\phi_k(\r_1;t)\phi_q(\r_2;t)$,
from which the variance of a many-particle operator $\hat A=\sum_j \hat a(\r)$ is computed,
\beqn\label{VAR_GEN}
& & \frac{1}{N} \Delta^2_{\hat A}(t) = 
\frac{1}{N}\left(\langle\Psi(t)|\hat A^2|\Psi(t)\rangle-\langle\Psi(t)|\hat A|\Psi(t)\rangle^2\right) = \nonumber \\
& & \quad = \frac{1}{N}\Bigg\{\sum_j n_j(t) \int d\r \phi_j^\ast(\r;t) \hat a^2(\r) \phi_j(\r;t) -
\left[\sum_j n_j(t) \int d\r \phi_j^\ast(\r;t) \hat a(\r) \phi_j(\r;t)\right]^2 + \nonumber \\
& & \quad + \sum_{jpkq} \rho_{jpkq}(t)\left[\int d\r \phi_j^\ast(\r;t) \hat a(\r) \phi_k(\r;t)\right]
\left[\int d\r \phi_p^\ast(\r;t) \hat a(\r) \phi_q(\r;t)\right]\Bigg\}. \
\eeqn
To compute the various terms for the position, momentum, and angular-momentum variance numerically
we work in coordinate representation and operate on orbitals
first with coordinate derivatives and then with coordinate multiplications.
Thus,
for the position operator $\hat a(\r)=\hat x$ and $\hat a^2(\r)=\hat x^2$ and likewise for $\hat a(\r)=\hat y$,
for the momentum operator
$\hat a(\r)=\frac{1}{i}\frac{\partial}{\partial x}$ and $\hat a^2(\r)=-\frac{\partial^2}{\partial x^2}$
and likewise for $\hat a(\r)=\frac{1}{i}\frac{\partial}{\partial y}$,
and for the angular-momentum operator
$\hat a(\r)=\frac{1}{i}\left(x \frac{\partial}{\partial y} - y \frac{\partial}{\partial x}\right)$
and
$\hat a^2(\r) = -x^2 \frac{\partial^2}{\partial y^2} -y^2 \frac{\partial^2}{\partial x^2}
+ 2 yx \frac{\partial}{\partial y}\frac{\partial}{\partial x} + x\frac{\partial}{\partial x}+y\frac{\partial}{\partial y}$.
For the numerical solution we use a grid of $64^2$ points in a box of size $[-8,8) \times [-8,-8)$ with periodic boundary conditions.
Convergence of the results with respect to the
number of grid points has been checked using a grid of $128^2$ points.

\begin{figure}[!]
\begin{center}
\hglue -1.0 truecm
\includegraphics[width=0.345\columnwidth,angle=-90]{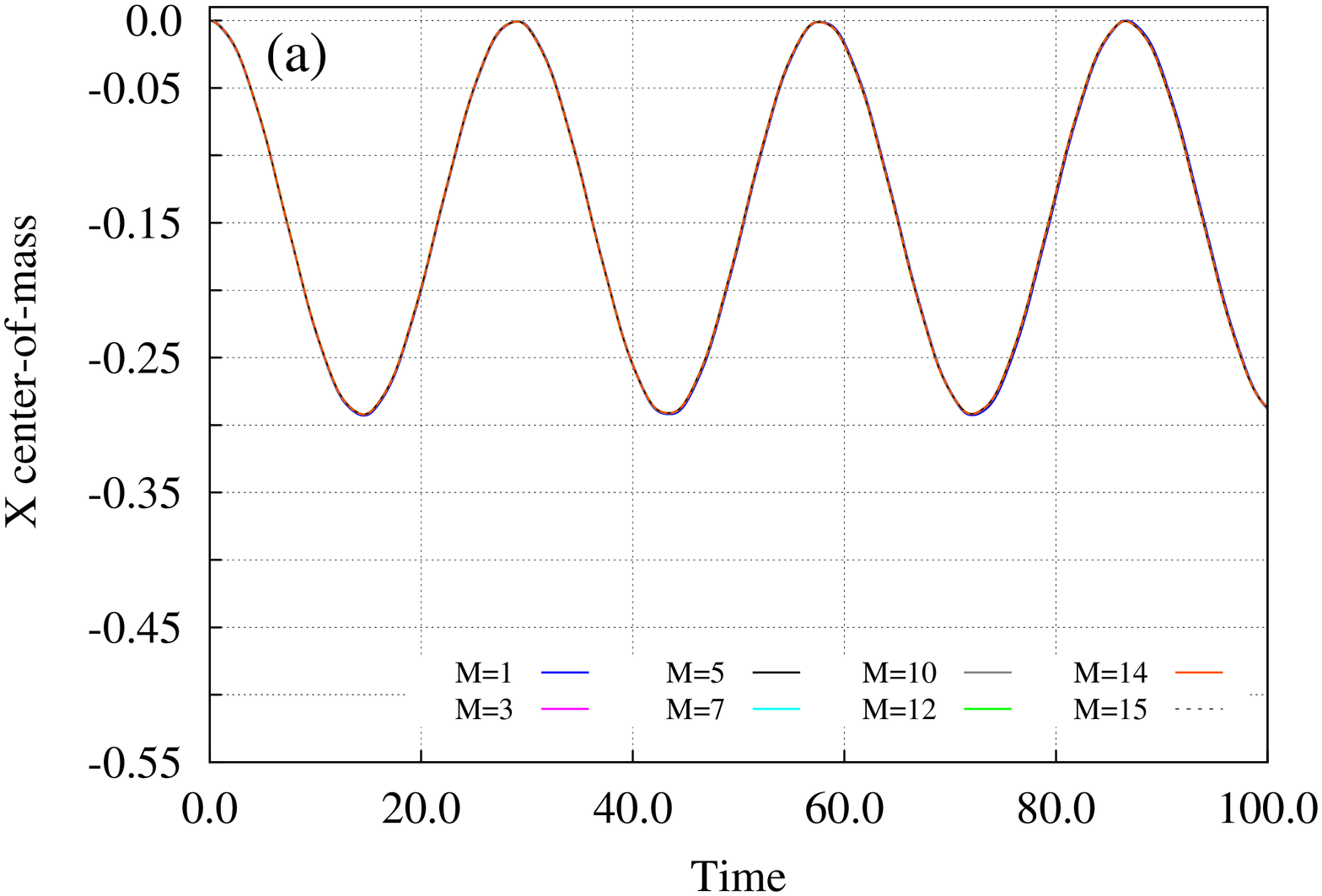}
\includegraphics[width=0.345\columnwidth,angle=-90]{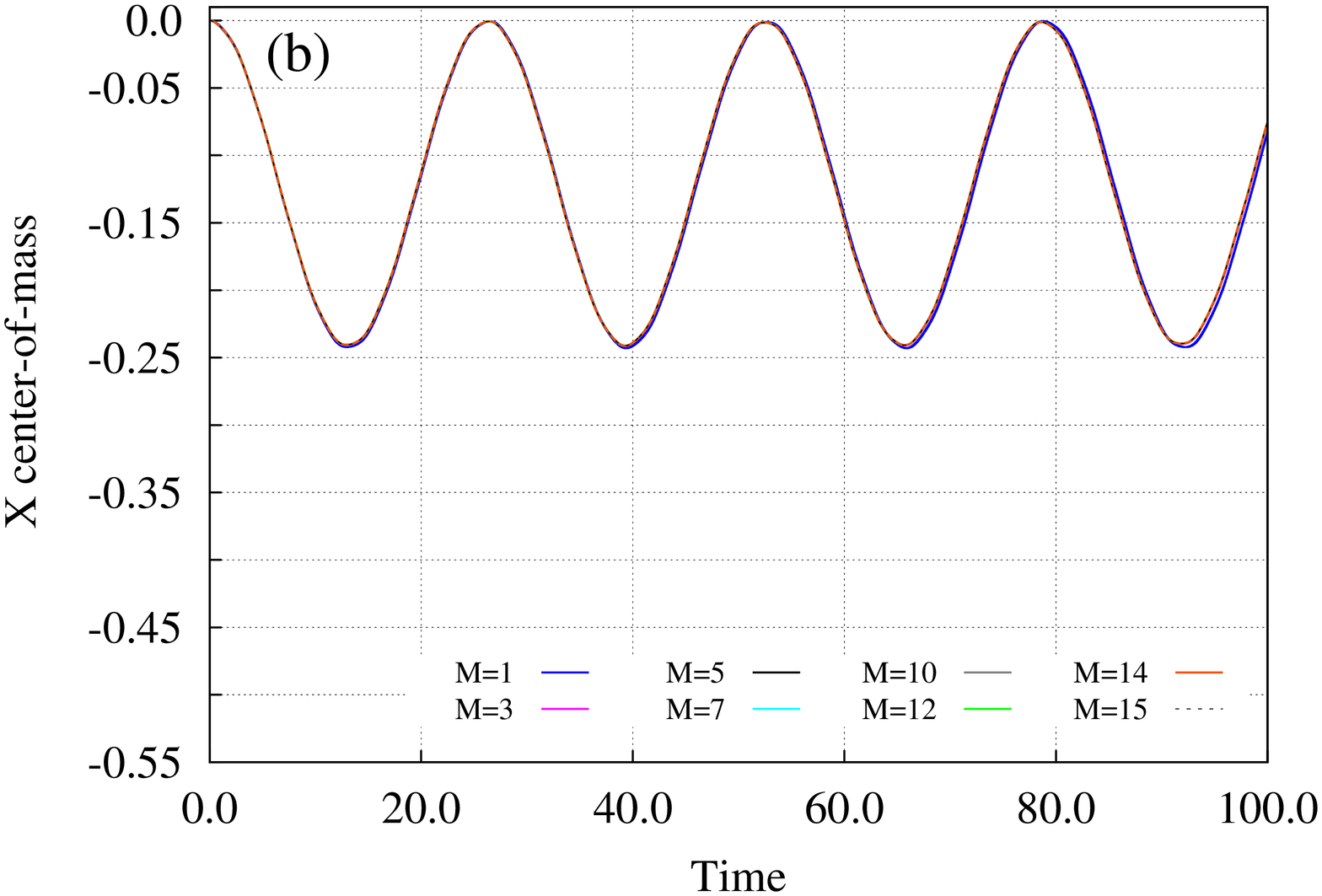}
\hglue -1.0 truecm
\includegraphics[width=0.345\columnwidth,angle=-90]{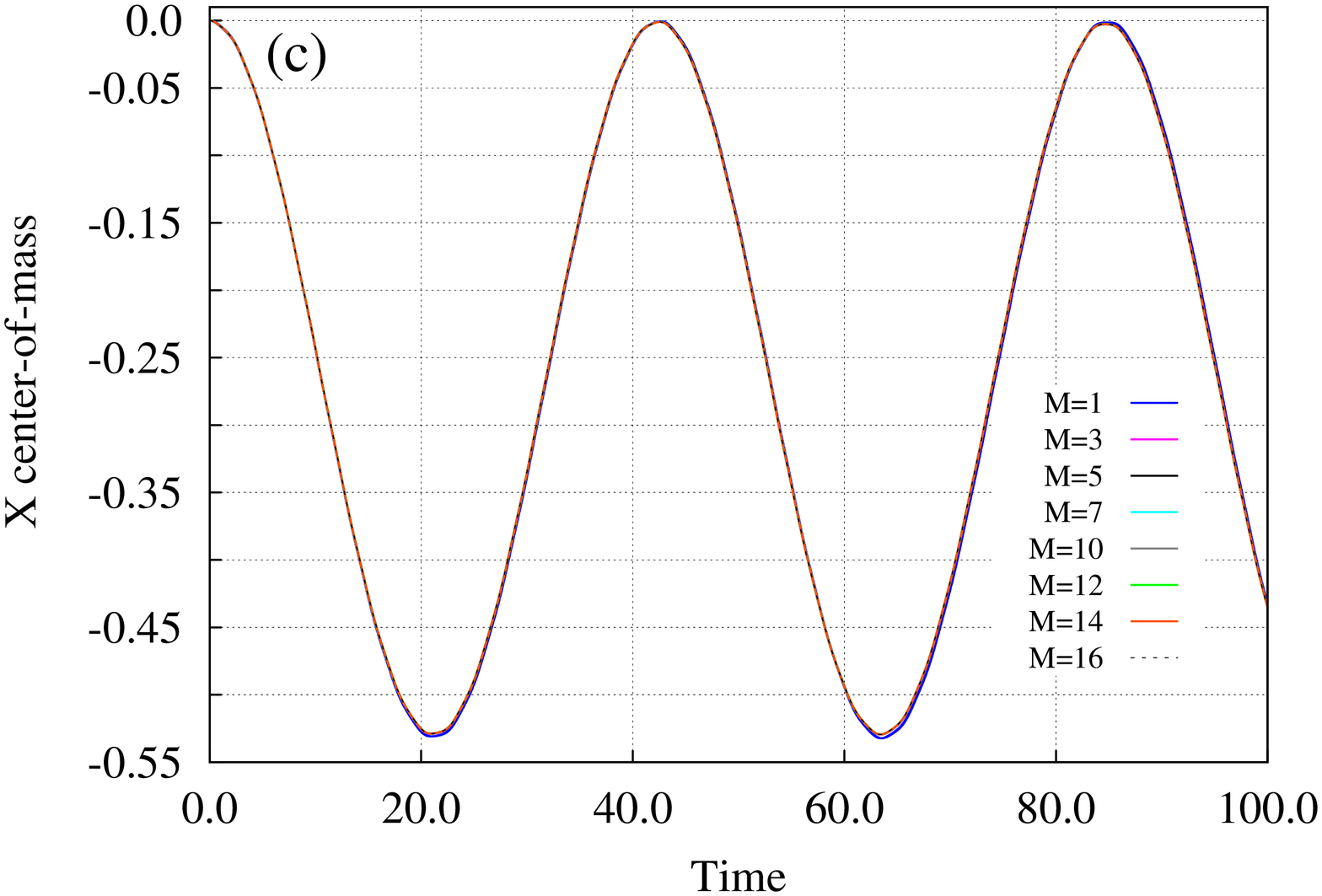}
\includegraphics[width=0.345\columnwidth,angle=-90]{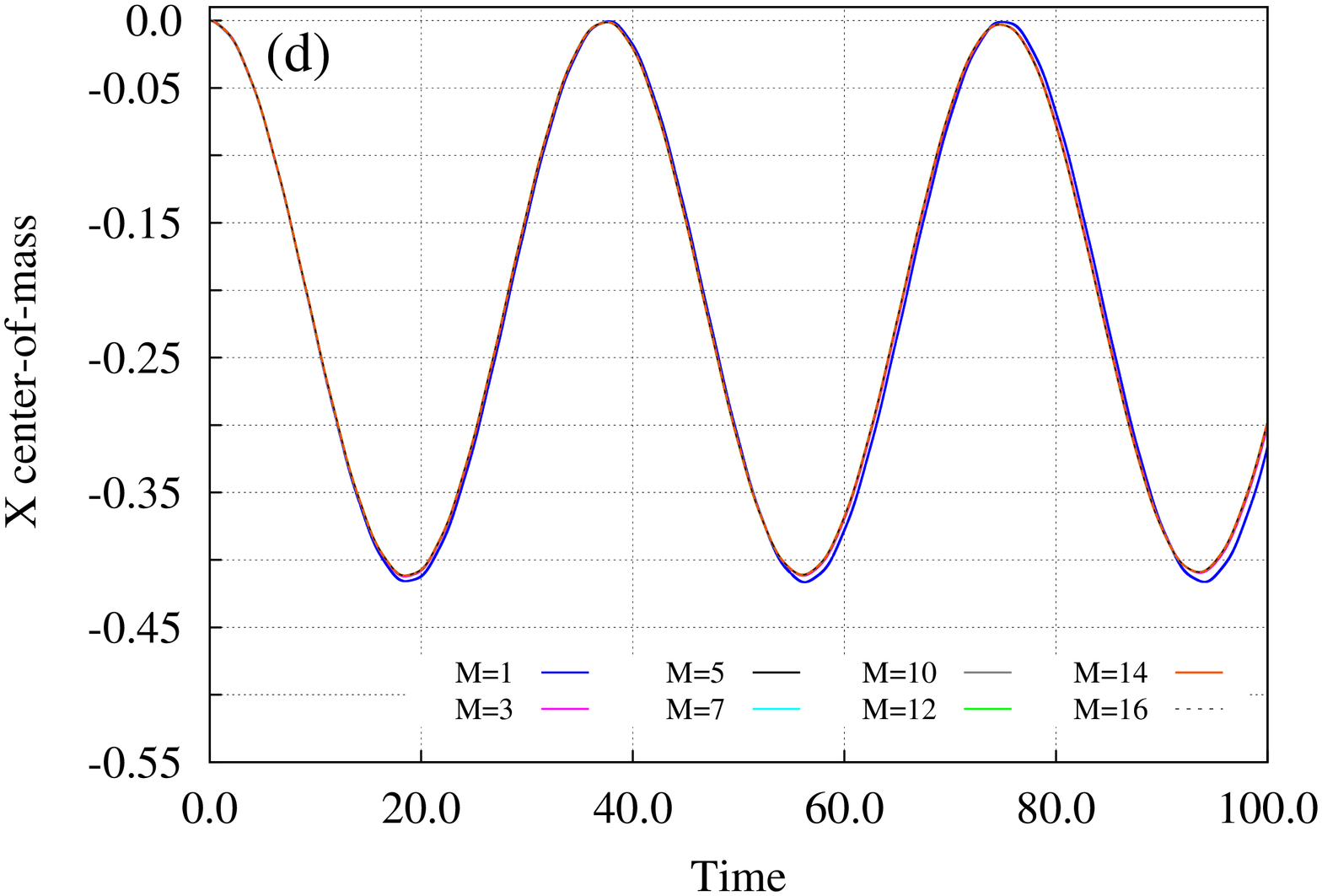}
\end{center}
\vglue 0.75 truecm
\caption{Center-of-mass dynamics following a potential quench.
The mean-field ($M=1$ time-adaptive orbitals) and many-body (using $M=3$, $5$, $7$, $10$, $12$, $14$, and $15$, $16$ time-adaptive orbitals)
time-dependent expectation value of the center-of-mass, $\frac{1}{N}\langle\Psi|\hat X|\Psi\rangle(t)$,
of $N=10$ bosons in the annuli with barrier heights and interaction strengths:
(a) $V_0=5$, $\lambda_0=0.02$;
(b) $V_0=5$, $\lambda_0=0.04$;
(c) $V_0=10$, $\lambda_0=0.02$; and 
(d) $V_0=10$, $\lambda_0=0.04$
following a sudden potential tilt by $0.01 x$.
The respective depletions are plotted in Fig.~\ref{f2}
and the position, momentum, and angular-momentum
variances in Figs.~\ref{f3}, \ref{f4}, and \ref{f5}, respectively.
See the text for more details.
The quantities shown are dimensionless.}
\label{f1}
\end{figure}

We begin with the dynamics of $N=10$ bosons in the annulus.
Following the sudden tilt of the potential,
the bosons start to flow to the left.
To quantify their sloshing dynamics,
Fig.~\ref{f1} shows the time-dependent center-of-mass,
$\frac{1}{N}\langle\Psi|\hat X|\Psi\rangle(t)$,
for the two barrier heights, $V_0=5$ and $V_0=10$,
and the two interaction strengths, $\lambda_0=0.02$ and $\lambda_0=0.04$
[we mention that $\frac{1}{N}\langle\Psi|\hat Y|\Psi\rangle(t)=0$ due to the $Y \to -Y$ reflection symmetry].
The dynamics of $\frac{1}{N}\langle\Psi|\hat X|\Psi\rangle(t)$ appears to be almost periodic and rather simple.
We examine the amplitude and frequency of oscillations.
It is useful to compare the amplitude of
the center-of-mass motion with the radius of the (un-tilted) annulus.
The radius of the density at its maximal value, $R$, is determined numerically
using a computation with a resolution of $256^2$ grid points as
$R=1.75(0)$ for $V_0=5, \lambda_0=0.02$, and
$R=2.06(2)$ for $V_0=10, \lambda_0=0.02$ \cite{var2}.
From Fig.~\ref{f1} we see that the amplitude is about $13\%$-$25\%$ of the radius,
implying a mild sloshing of the density along the tilted annulus.
The amplitude increases with the radius of the annulus and decreases with the interaction strength,
where the latter implies that it is more difficult to compress the BEC for a stronger interaction. 
The decrease of the frequency of oscillations with $R$ ($V_0$) and increase with $\lambda_0$
are compatible with angular excitations, also see \cite{var2}.
Last but not least, convergence with $M$ is clearly seen.
In fact, here, already $M=1$ orbitals accurately describe the center-of-mass dynamics
(for short and intermediate times, and $M=3$ orbitals for all times). 

\begin{figure}[!]
\begin{center}
\hglue -1.0 truecm
\includegraphics[width=0.345\columnwidth,angle=-90]{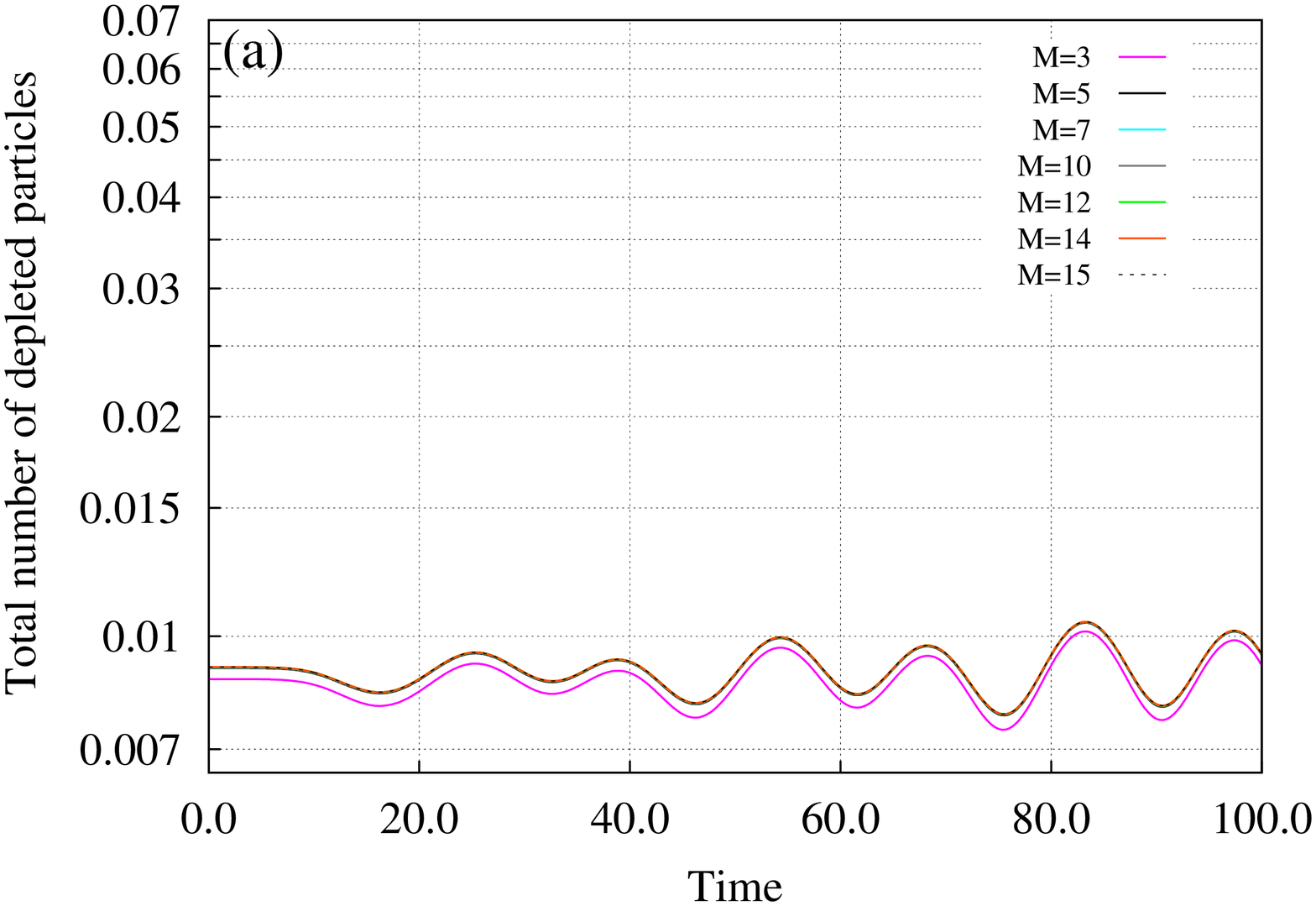}
\includegraphics[width=0.345\columnwidth,angle=-90]{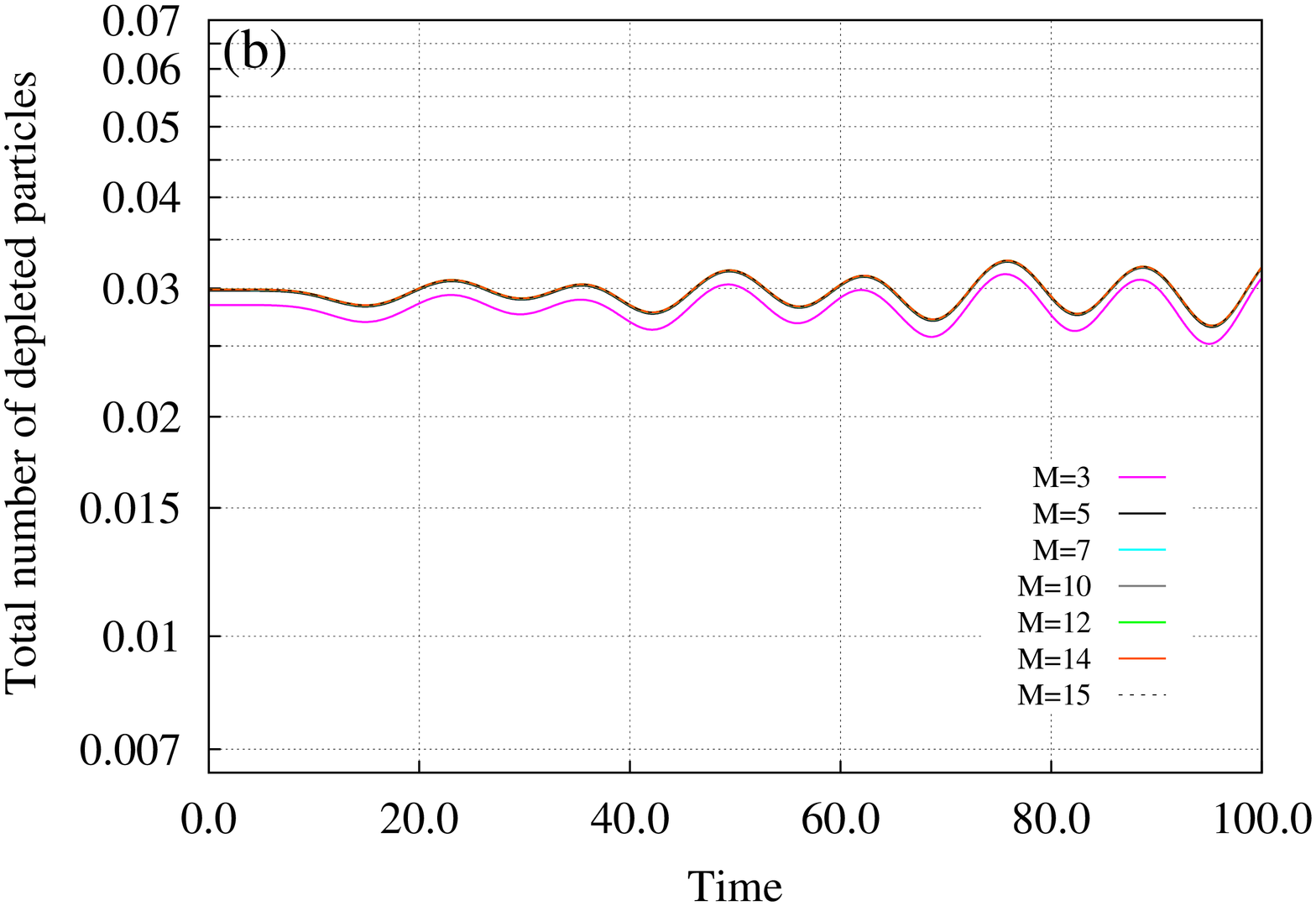}
\hglue -1.0 truecm
\includegraphics[width=0.345\columnwidth,angle=-90]{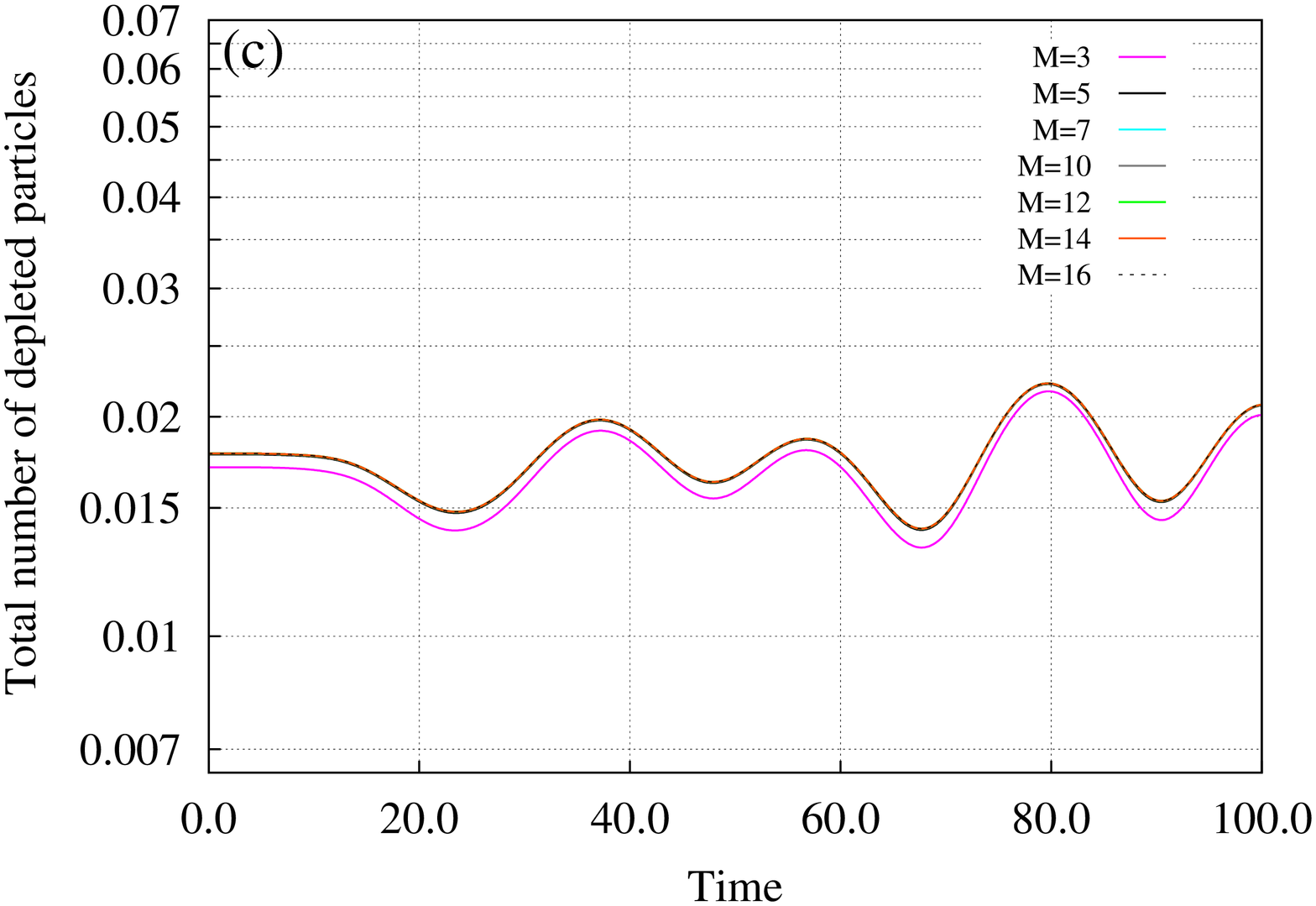}
\includegraphics[width=0.345\columnwidth,angle=-90]{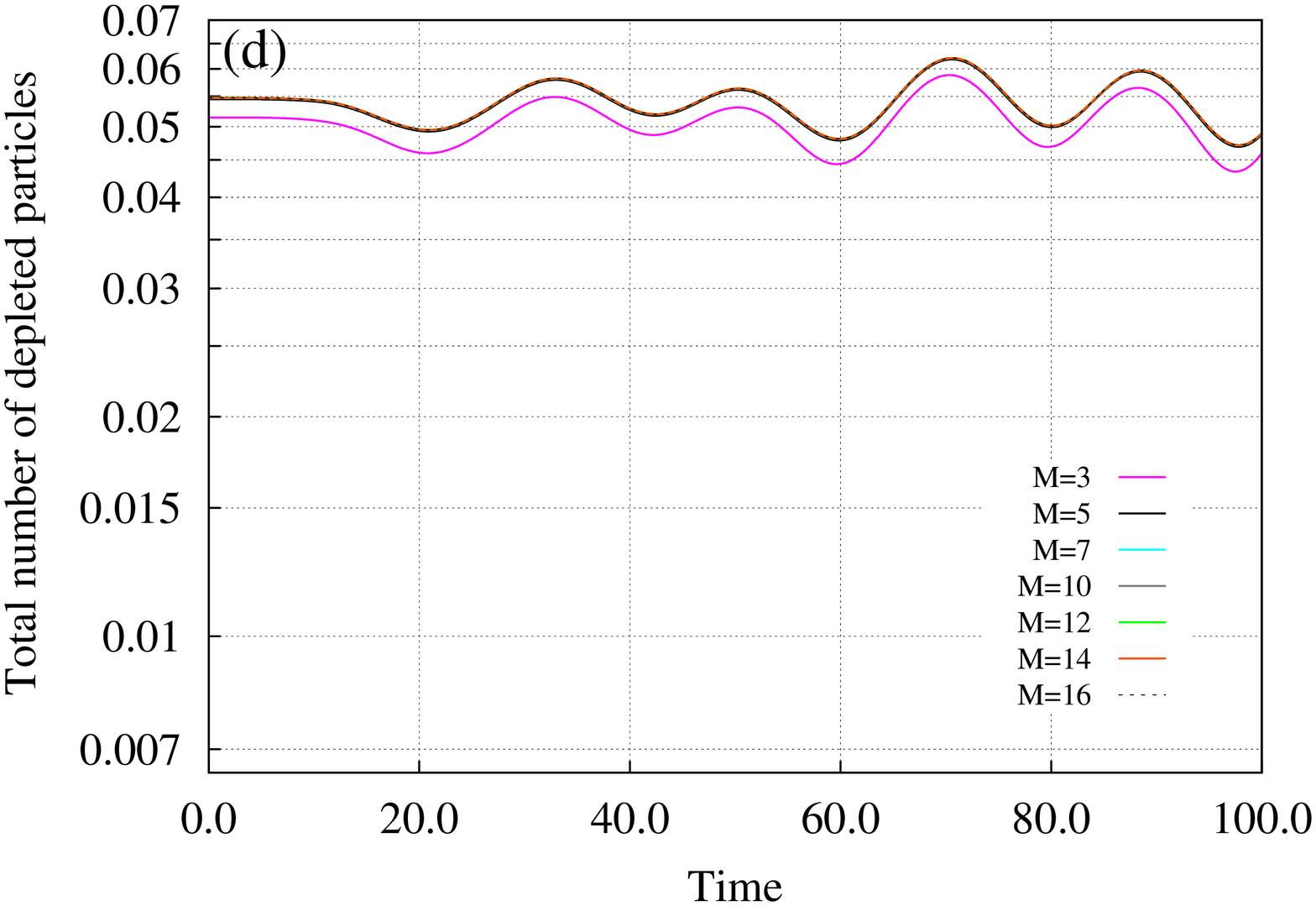}
\end{center}
\vglue 0.75 truecm
\caption{Depletion dynamics following a potential quench.
The time-dependent total number of depleted particles, $N-n_1(t)$,
of $N=10$ bosons following a sudden potential tilt by $0.01 x$
for annuli with barrier heights and interaction strengths
(a) $V_0=5$, $\lambda_0=0.02$;
(b) $V_0=5$, $\lambda_0=0.04$;
(c) $V_0=10$, $\lambda_0=0.02$; and 
(d) $V_0=10$, $\lambda_0=0.04$.
$M=3$, $5$, $7$, $10$, $12$, $14$, and $15$, $16$ time-adaptive orbitals are used.
The respective position, momentum, and angular-momentum
variances are plotted in Figs.~\ref{f3}, \ref{f4}, and \ref{f5}.
See the text for more details.
The quantities shown are dimensionless.
}
\label{f2}
\end{figure}

Fig.~\ref{f2} depicts the total number of depleted particles,
$N-n_1(t)$, out of $N=10$ bosons
in the tilted annulus.
During the dynamics, the depletion is rather small,
ranging from less than $0.012$ of a particle out of $N=10$ particles ($0.12\%$) for $V_0=5, \lambda_0=0.02$
to less than $0.065$ of a particle out of $N=10$ particles ($0.65\%$) for $V_0=10, \lambda_0=0.04$.
Generally, the depletion increases with the annulus radius and interacting strength,
implying angular excitations, see \cite{var2}.
Finally, convergence with $M$ is clearly seen.
Now, $M=3$ orbitals nicely follow and $M=5$ orbitals
accurately describe the depletion dynamics,
see Fig.~\ref{f2}.
The small amount of time-dependent depletion
is in line with
the accurate description of the center-of-mass dynamics
by $M=1$ time adaptive orbitals, see Fig.~\ref{f1}.
Let us continue to the variances.

\begin{figure}[!]
\begin{center}
\vglue -1.25 truecm
\hglue -1.0 truecm
\includegraphics[width=0.2930\columnwidth,angle=-90]{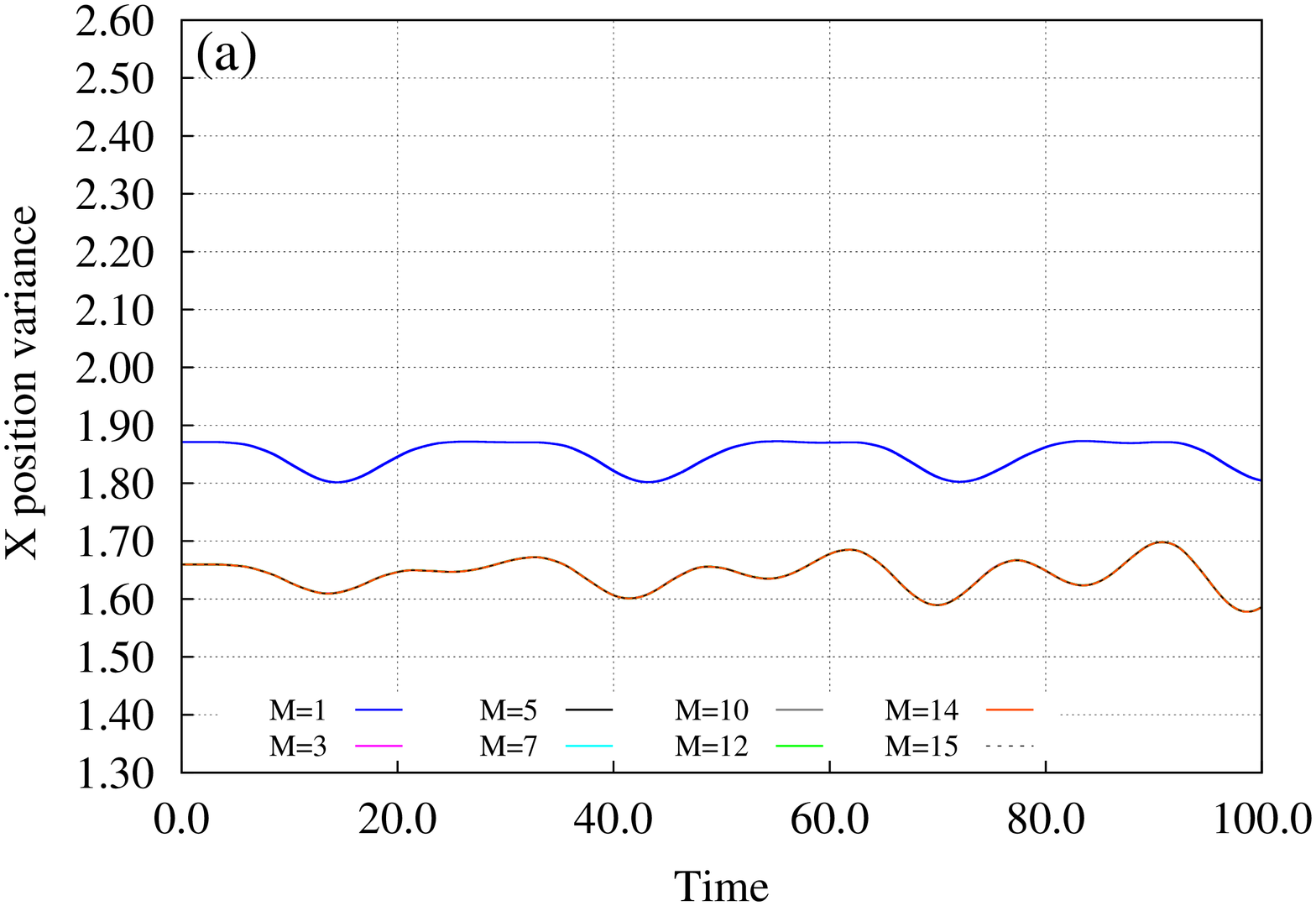}
\includegraphics[width=0.2930\columnwidth,angle=-90]{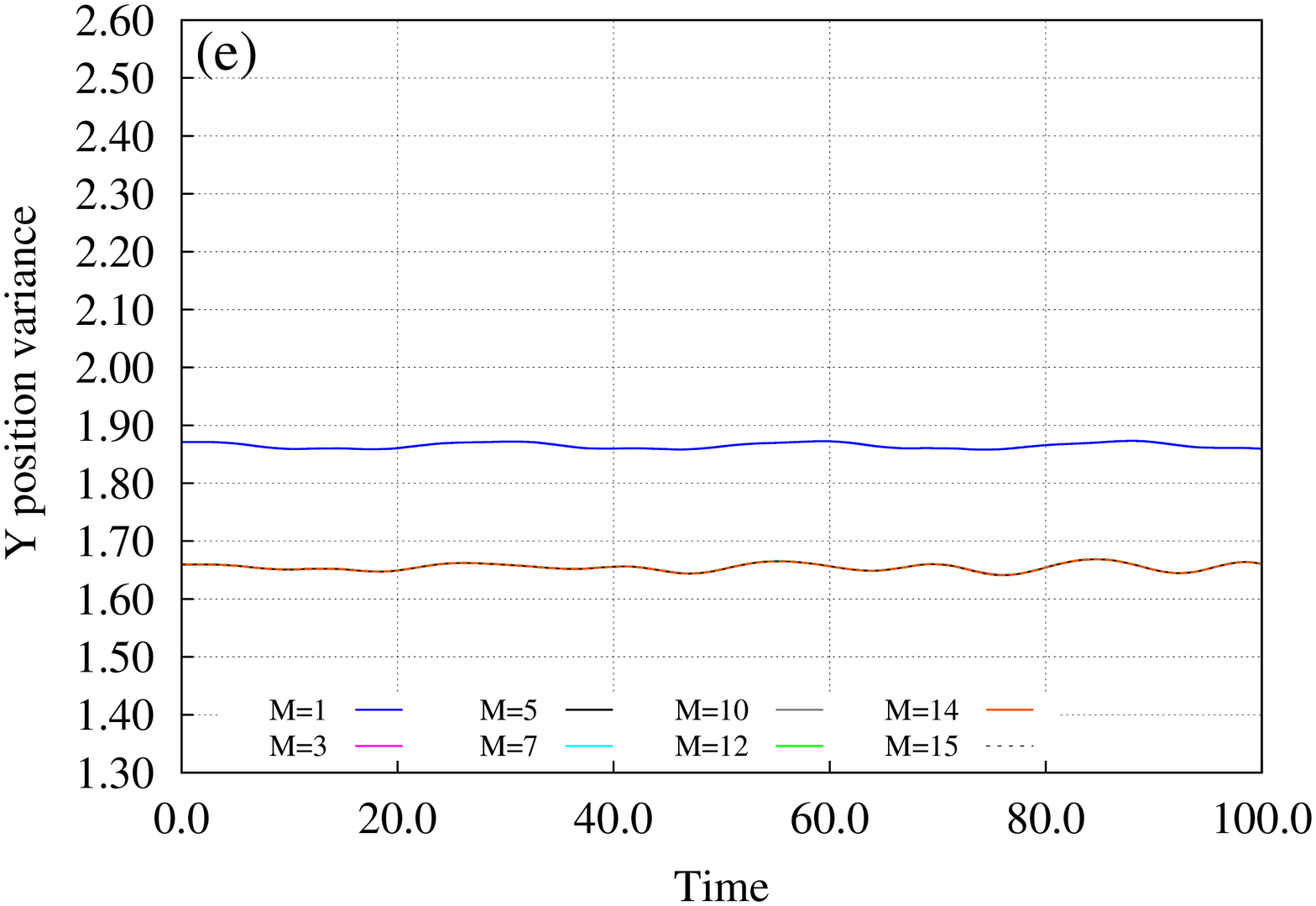}
\vglue 0.25 truecm
\hglue -1.0 truecm
\includegraphics[width=0.2930\columnwidth,angle=-90]{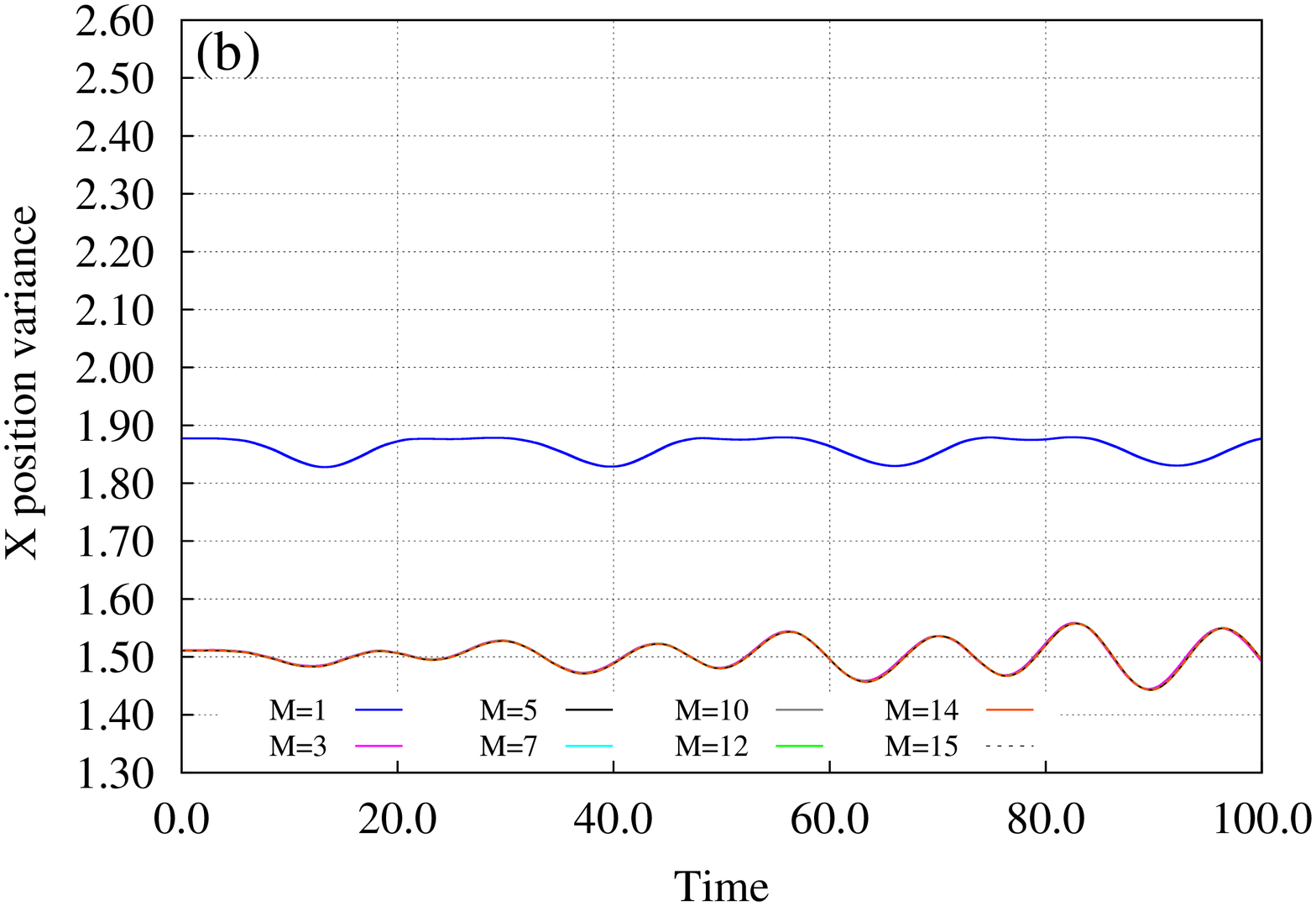}
\includegraphics[width=0.2930\columnwidth,angle=-90]{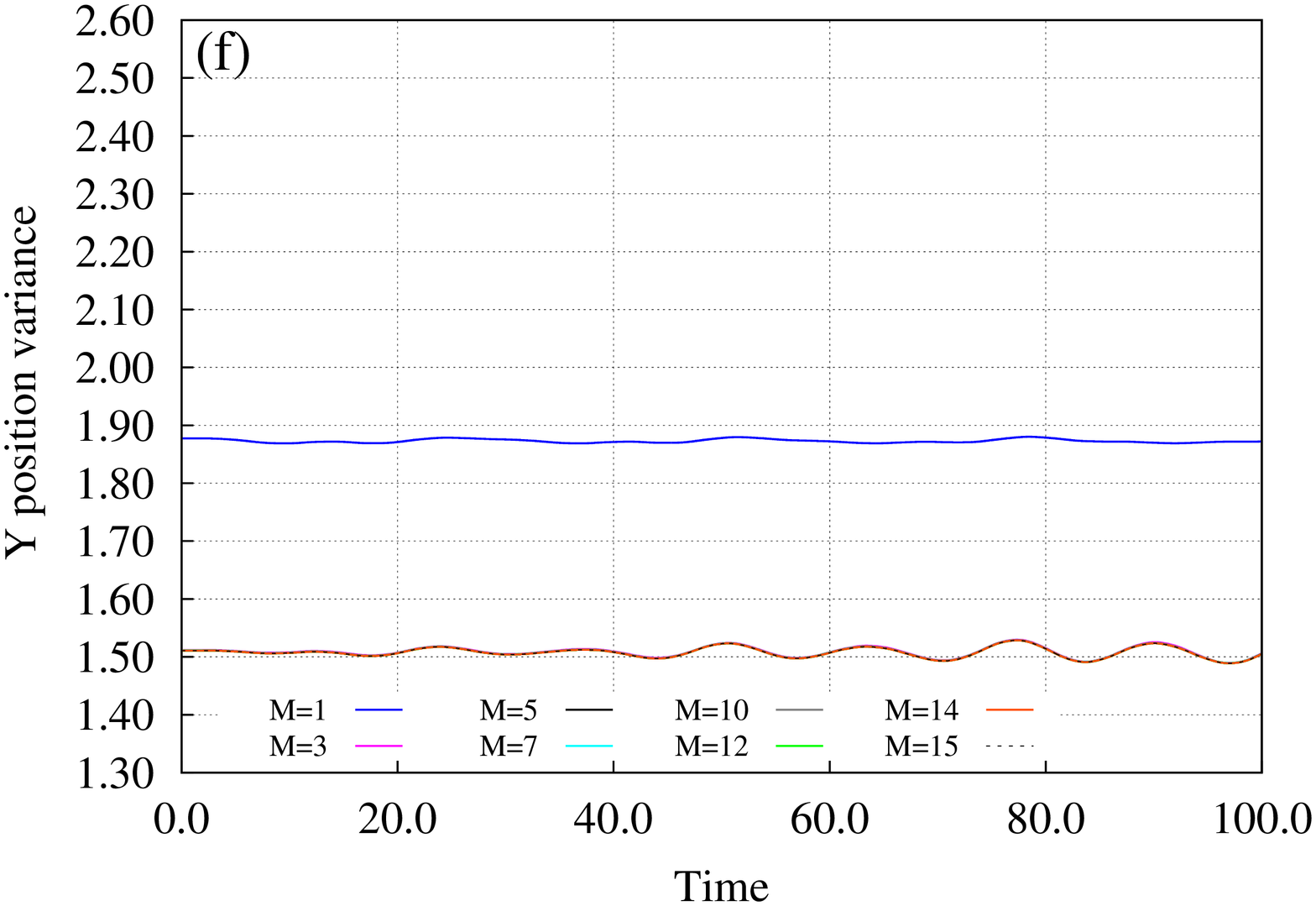}
\vglue 0.25 truecm
\hglue -1.0 truecm
\includegraphics[width=0.2930\columnwidth,angle=-90]{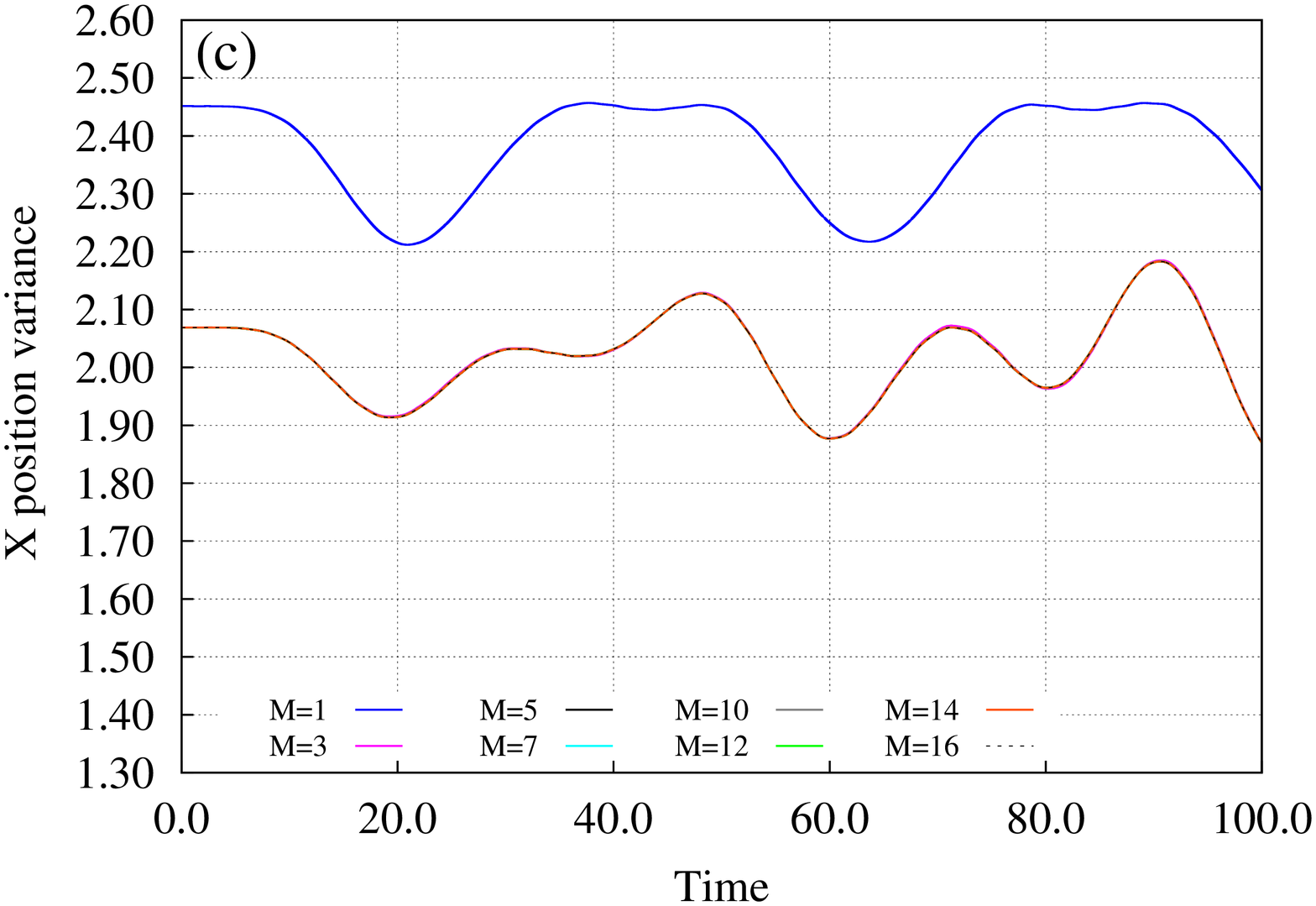}
\includegraphics[width=0.2930\columnwidth,angle=-90]{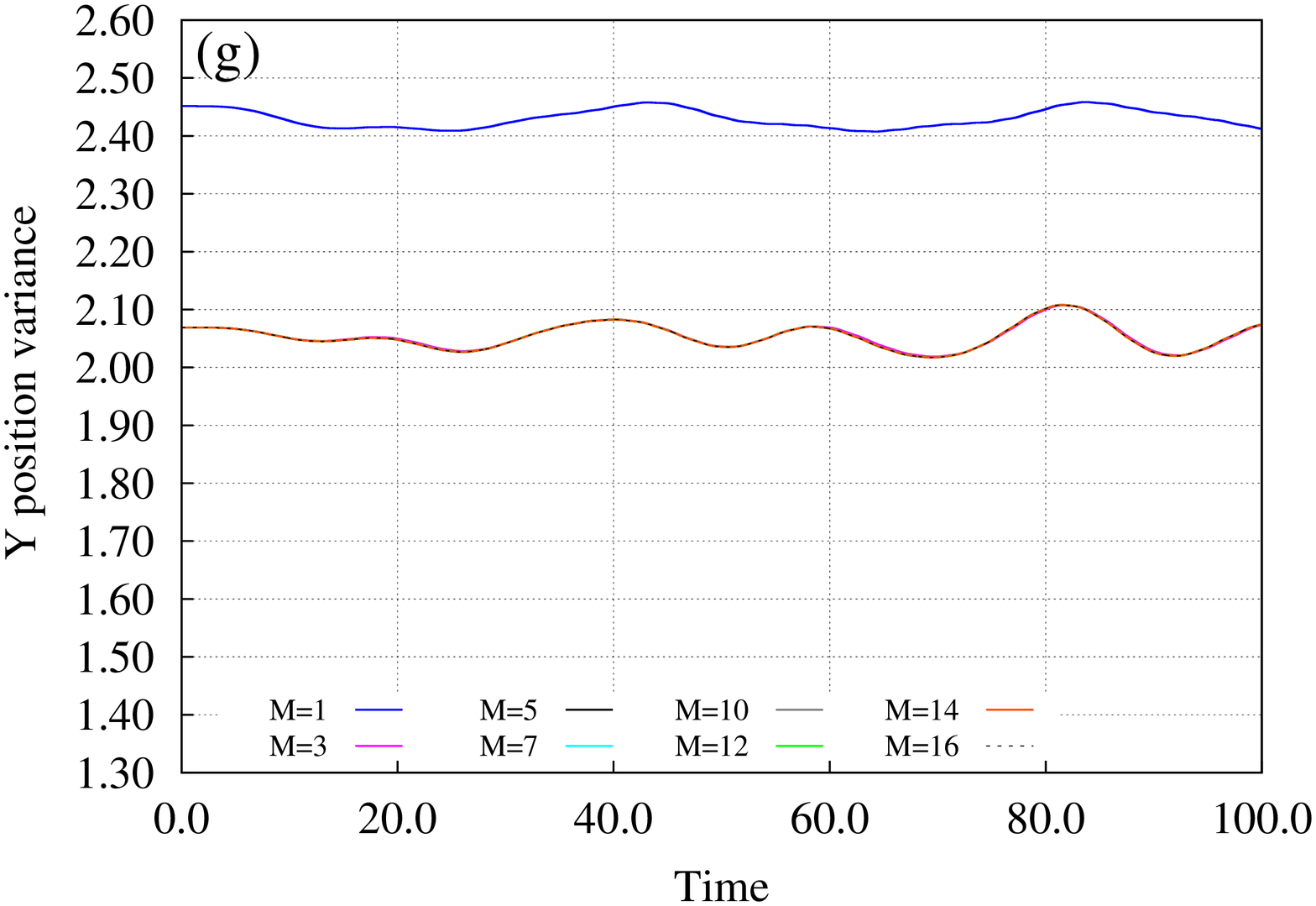}
\vglue 0.25 truecm
\hglue -1.0 truecm
\includegraphics[width=0.2930\columnwidth,angle=-90]{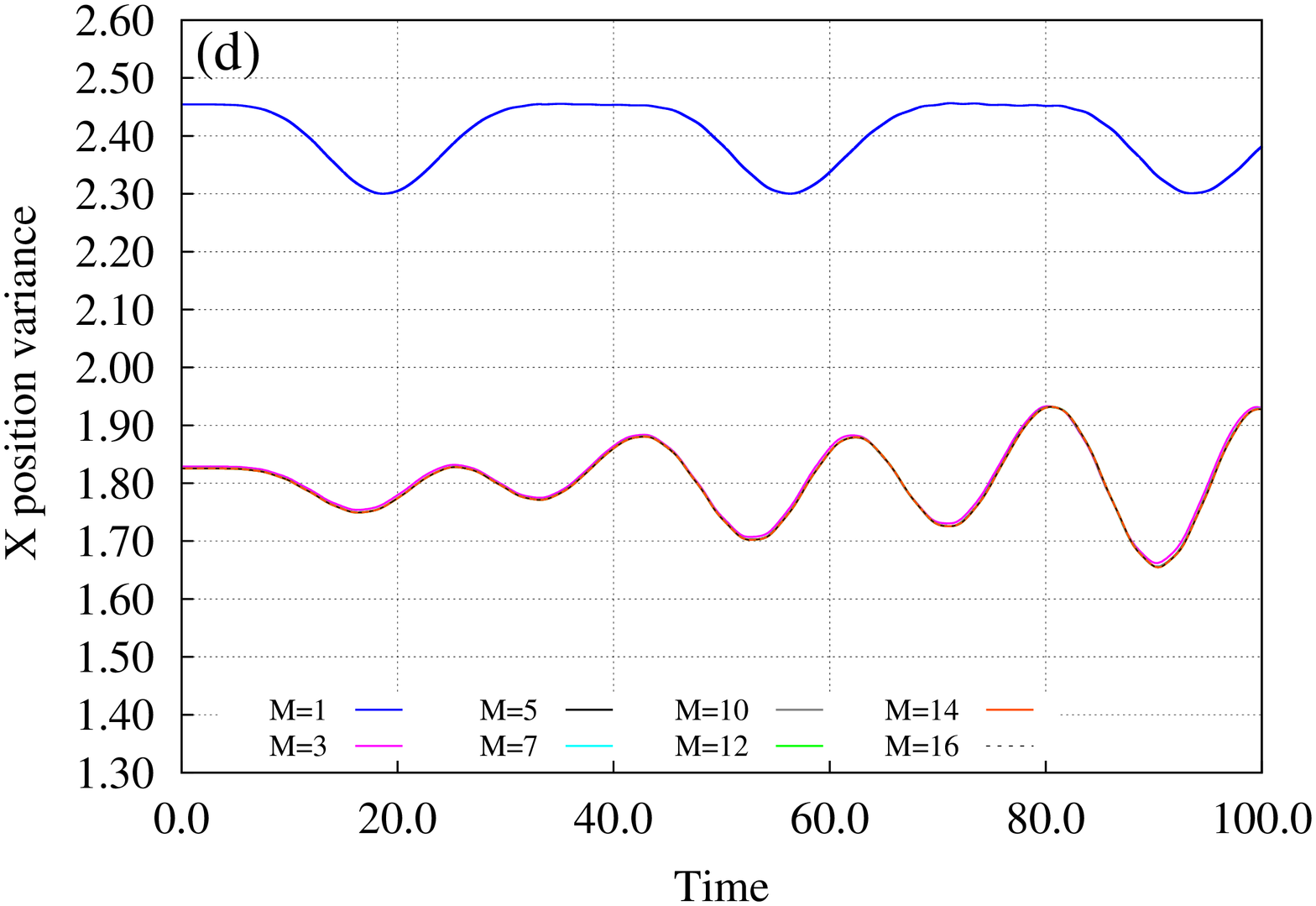}
\includegraphics[width=0.2930\columnwidth,angle=-90]{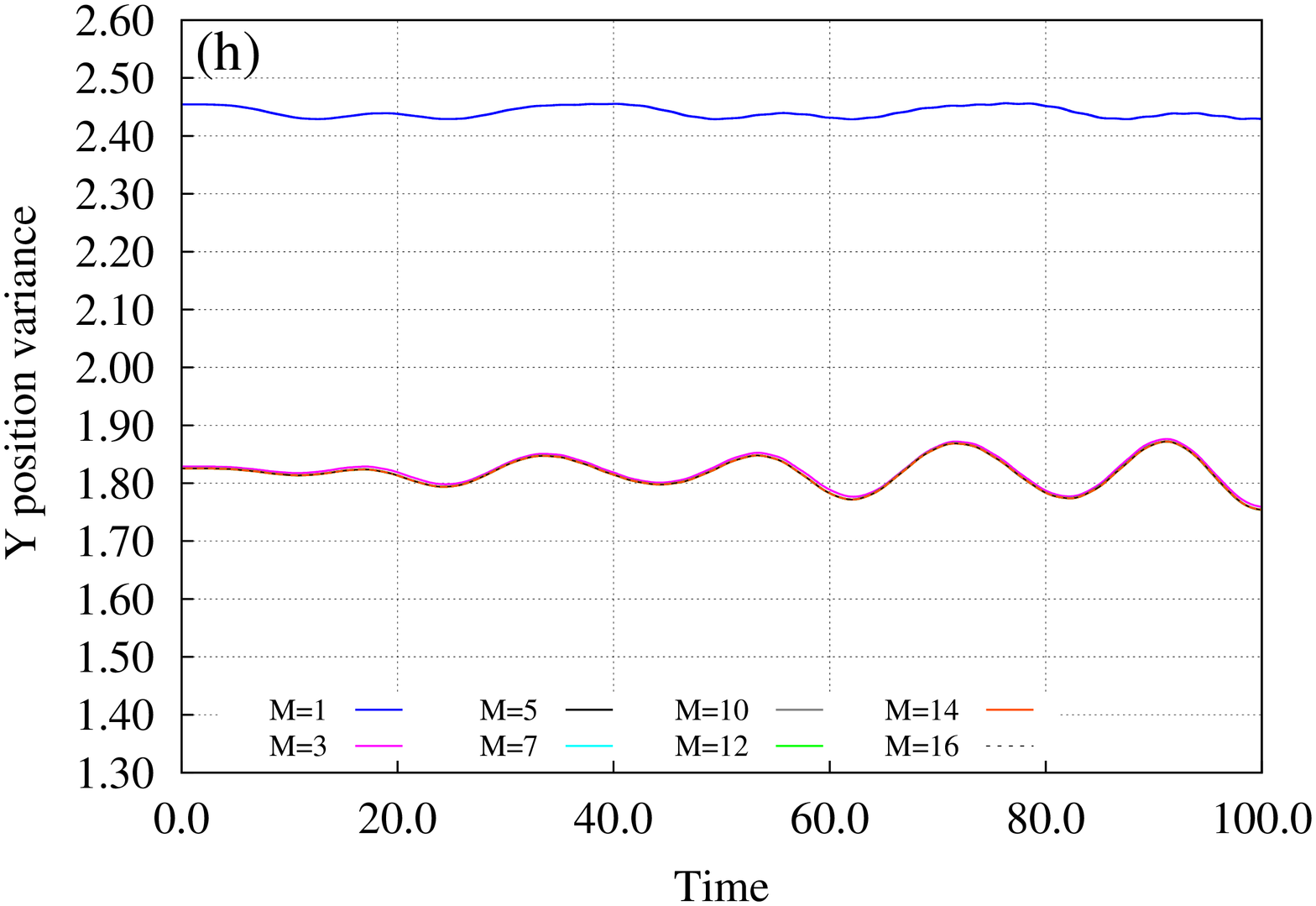}
\end{center}
\vglue 0.25 truecm
\caption{Position variance dynamics following a potential quench.
The mean-field ($M=1$ time-adaptive orbitals) and many-body (using $M=3$, $5$, $7$, $10$, $12$, $14$, and $15$, $16$ time-adaptive orbitals)
time-dependent position variances per particle,
$\frac{1}{N}\Delta^2_{\hat X}(t)$ [left column, panels (a), (b), (c), and (d)] and
$\frac{1}{N}\Delta^2_{\hat Y}(t)$ [right column, panels (e), (f), (g), and (h)],
of $N=10$ bosons in the annuli 
with barrier heights and interaction strengths
(a) and (d) $V_0=5$, $\lambda_0=0.02$;
(b) and (e) $V_0=5$, $\lambda_0=0.04$;
(c) and (f) $V_0=10$, $\lambda_0=0.02$; and 
(d) and (g) $V_0=10$, $\lambda_0=0.04$
following a sudden potential tilt by $0.01 x$.
The respective depletions are plotted in Fig.~\ref{f2}.
See the text for more details.
The quantities shown are dimensionless.
}
\label{f3}
\end{figure}

Fig.~\ref{f3} plots the time-dependent
many-particle position variance per particle, $\frac{1}{N} \Delta^2_{\hat X}(t)$ and $\frac{1}{N} \Delta^2_{\hat Y}(t)$,
for the two barrier heights and two interaction strengths.
There are several features that immediately are seen.
First, since rotational symmetry is lifted,
the dynamics of respective quantities along the x-axis and y-axis are different
[note that at $t=0$ the variances
$\frac{1}{N} \Delta^2_{\hat X}=\frac{1}{N} \Delta^2_{\hat Y}$
because the initial condition is the ground state of the un-tilted, isotropic annulus].
The mean-field ($M=1$) and many-body ($M \ge 3$) values are clearly separated from each other,
and the former lie about $10\%$-$25\%$ above the latter depending on the repulsion strength and barrier height, also see \cite{INF6,var2}.
This is despite the small amount of depletion, see Fig.~\ref{f2}.
Furthermore, the many-body and mean-field variances do not cross each other, see Fig.~\ref{f3},
indicating that the dynamics is mild and sufficiently close to the ground state manifold of states
(compare to \cite{var1} with interaction-quench dynamics in a single trap).

The mean-field position variance accounts for the geometry of the annulus and shape of the density,
and weakly depends on the interaction strength.
The many-body position variance incorporates
the (small amount of) depletion and hence strongly depends on the interaction strength.
Both the mean-field and many-body variances oscillate with a relatively small amplitude,
albeit with a different frequencies' content,
see in this respect \cite{var_ap1}.
This amplitude slightly decreases with the repulsion strength,
which correlates with the dependence of the center-of-mass dynamics on the interaction strength,
see Fig.~\ref{f1}.
Moreover, the amplitude of oscillations of the y-axis variances is smaller than that of the x-axis variances,
since the sloshing dynamics is primarily along the $x$ direction.
Last but not least is the so-called opposite anisotropy of the (position) variance \cite{var1}.
During the dynamics, there can occur instances where 
$\frac{1}{N} \Delta^2_{\hat X} > \frac{1}{N} \Delta^2_{\hat Y}$ at the many-body level ($M\ge 3$)
whereas
$\frac{1}{N} \Delta^2_{\hat X} < \frac{1}{N} \Delta^2_{\hat Y}$ at the mean-field level ($M=1$)
[or, in principle, vice versa, i.e.,
$\frac{1}{N} \Delta^2_{\hat X} < \frac{1}{N} \Delta^2_{\hat Y}$ at the many-body level
whereas
$\frac{1}{N} \Delta^2_{\hat X} > \frac{1}{N} \Delta^2_{\hat Y}$ at the mean-field level].
Examples for the former
can be readily found for $V_0=10$, $\lambda_0=0.02$, see Fig.~\ref{f3}c,g around $t=70$,
and for $V_0=10$, $\lambda_0=0.04$, see Fig.~\ref{f3}d,h around $t=100$, 
signifying among others that correlations `win' over shape.
Finally, we see that already $M=3$ orbitals accurately describe the dynamics of the position variance.

\begin{figure}[!]
\begin{center}
\vglue -1.25 truecm
\hglue -1.0 truecm
\includegraphics[width=0.2930\columnwidth,angle=-90]{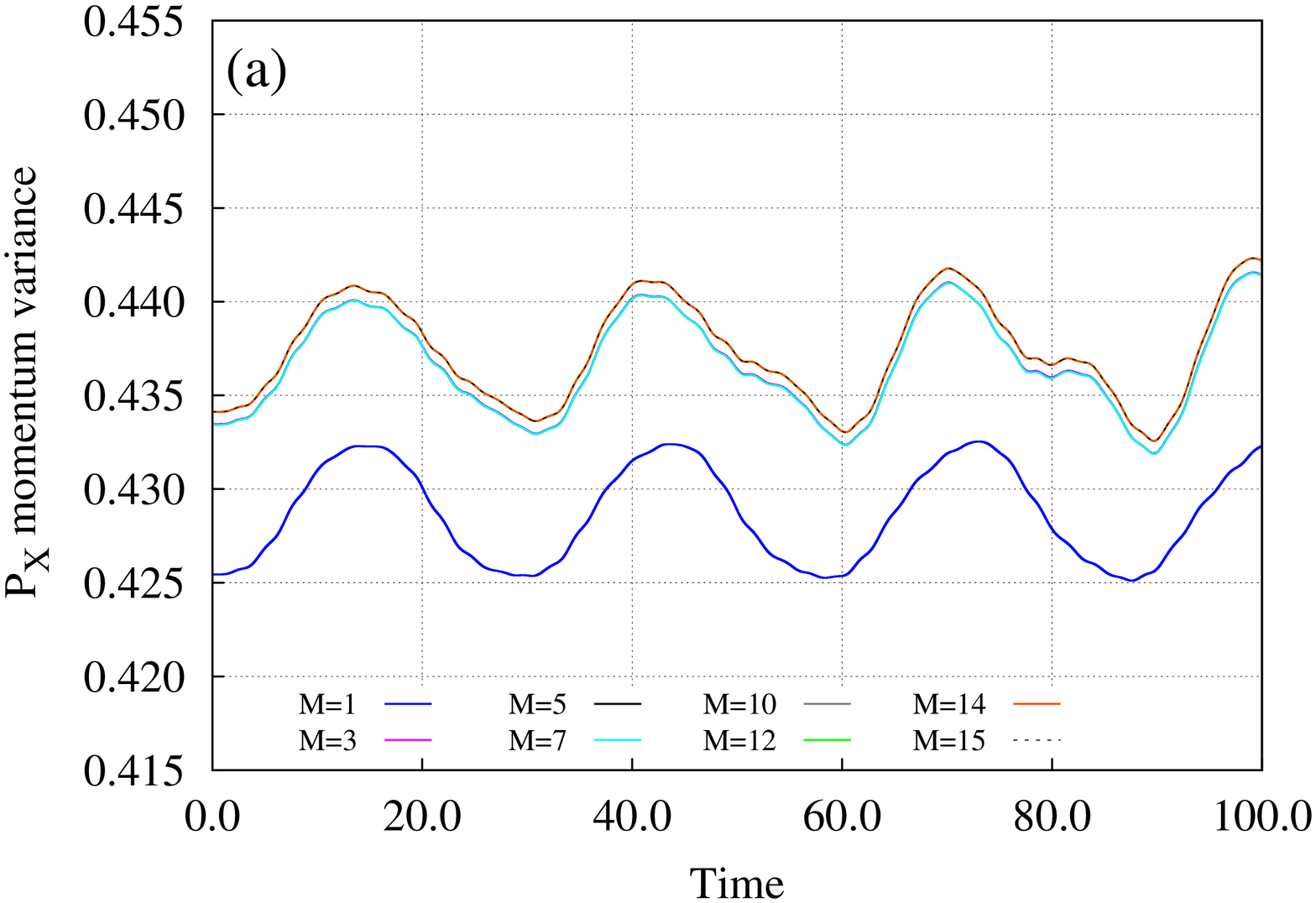}
\includegraphics[width=0.2930\columnwidth,angle=-90]{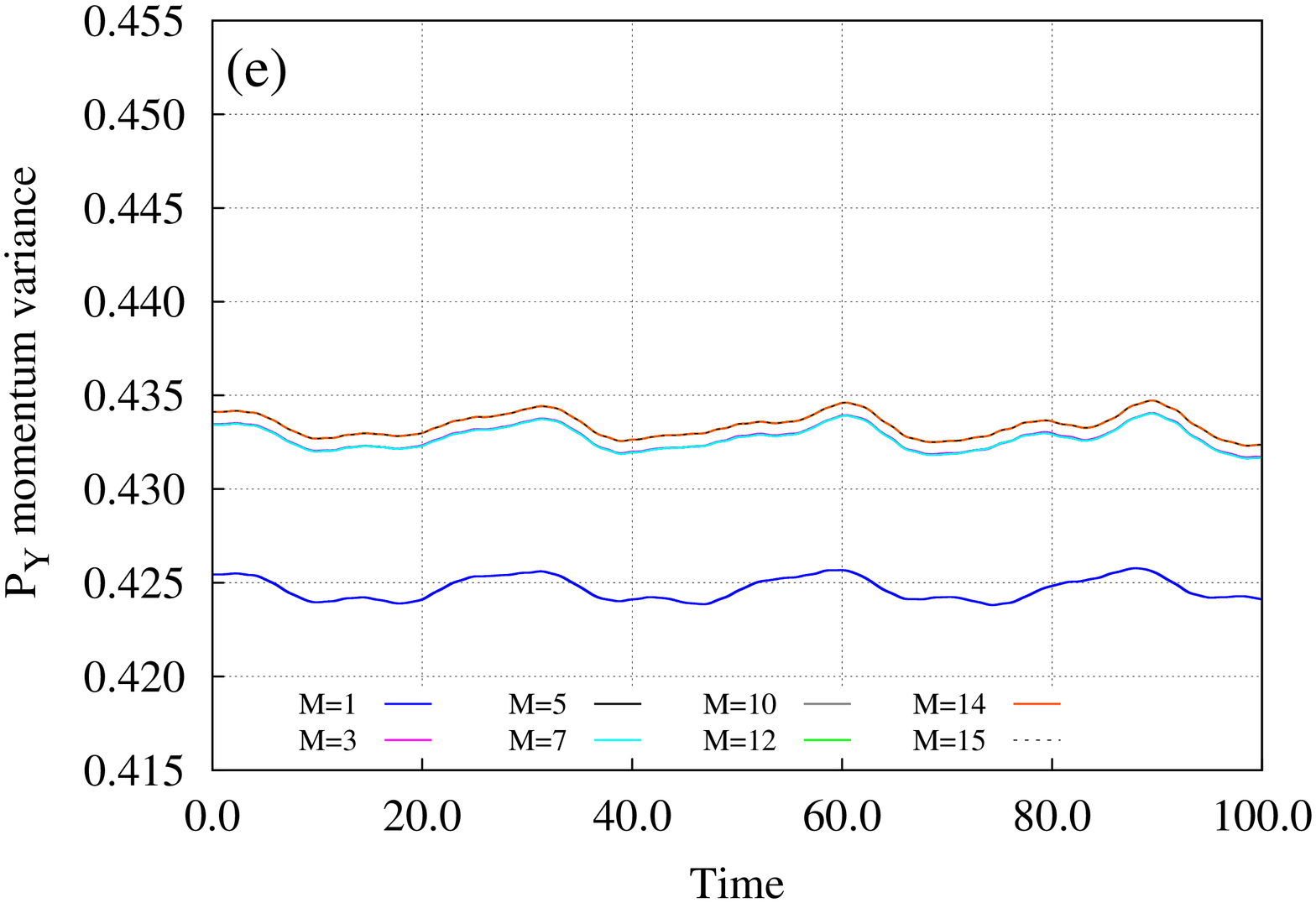}
\vglue 0.25 truecm
\hglue -1.0 truecm
\includegraphics[width=0.2930\columnwidth,angle=-90]{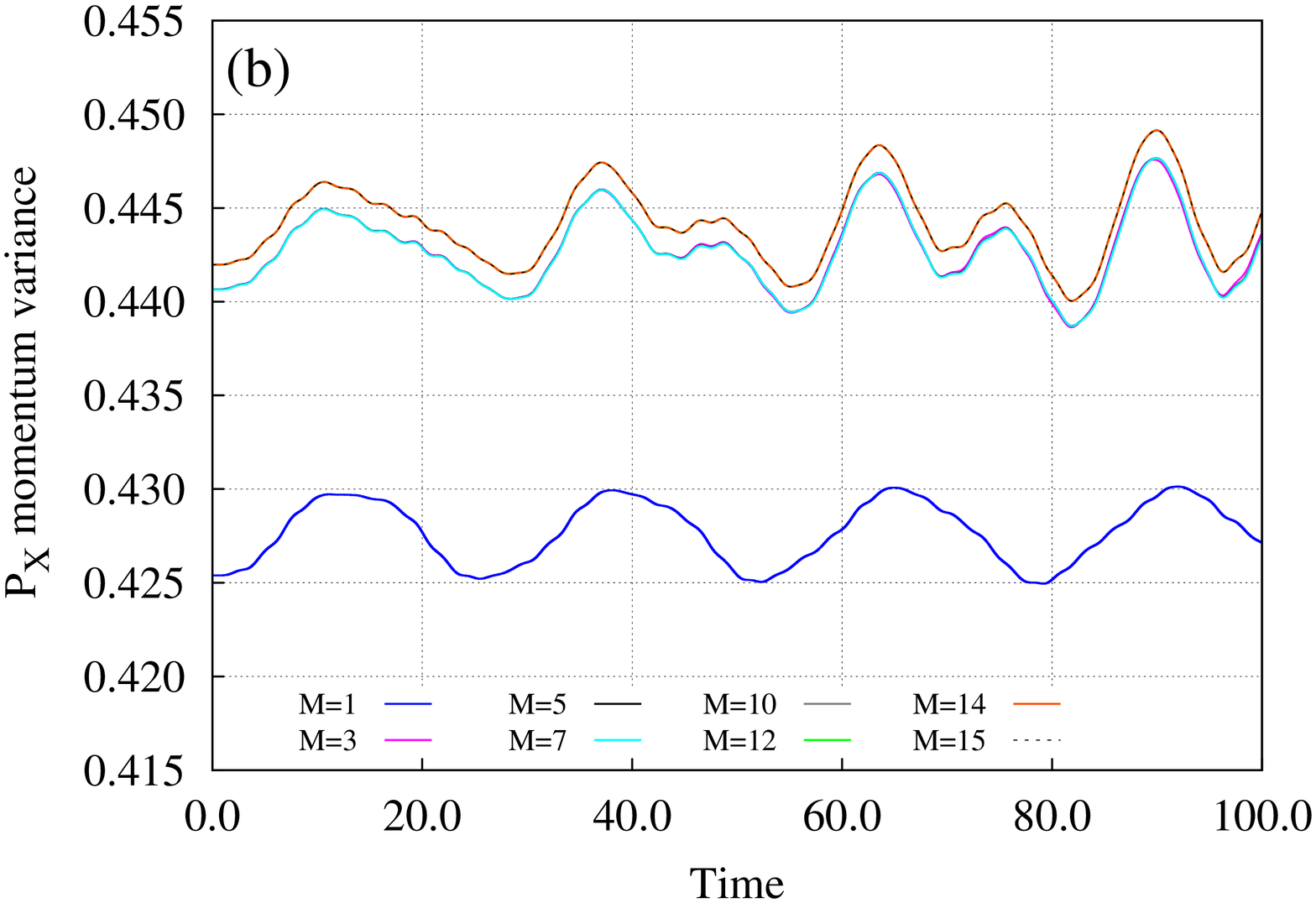}
\includegraphics[width=0.2930\columnwidth,angle=-90]{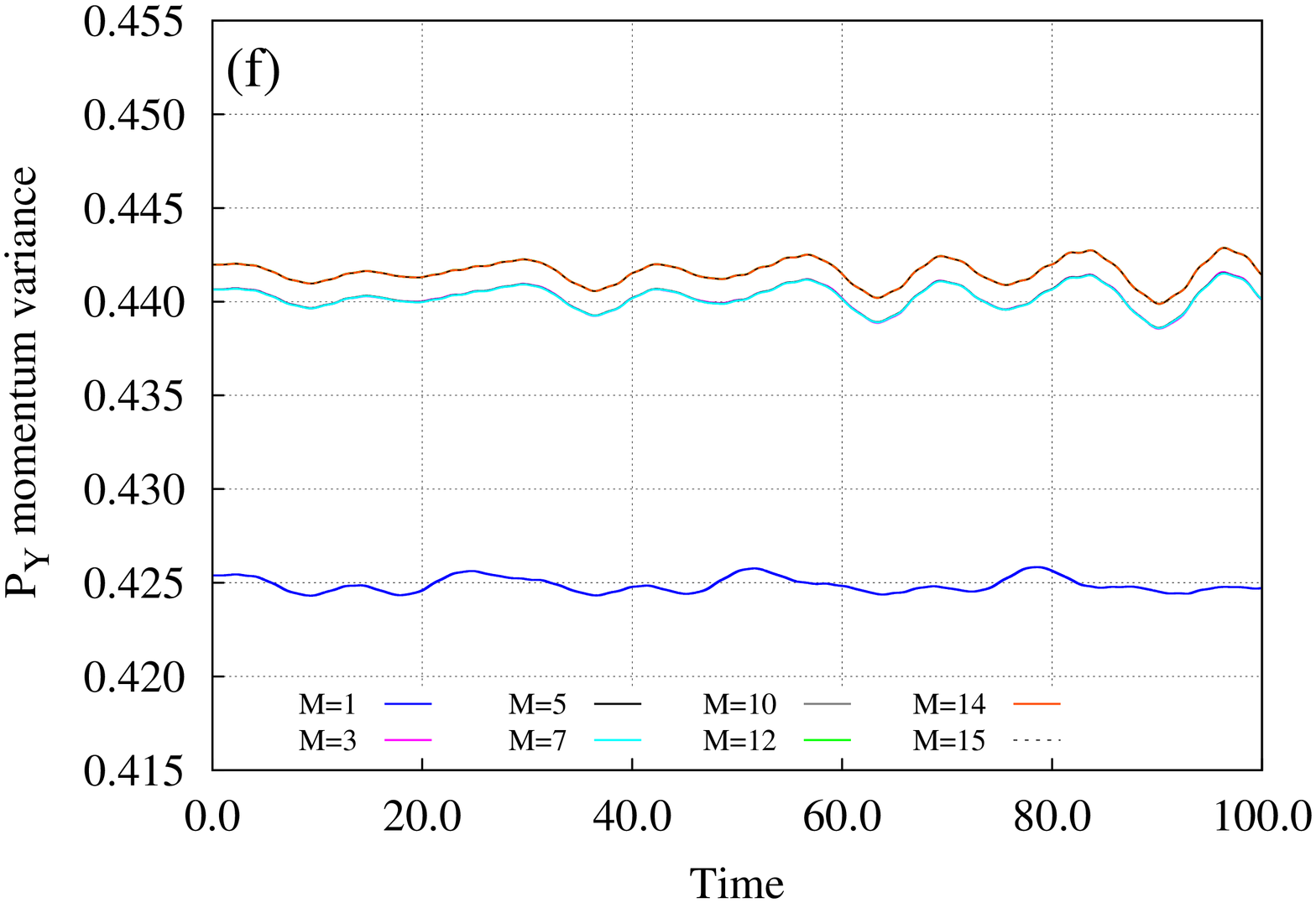}
\vglue 0.25 truecm
\hglue -1.0 truecm
\includegraphics[width=0.2930\columnwidth,angle=-90]{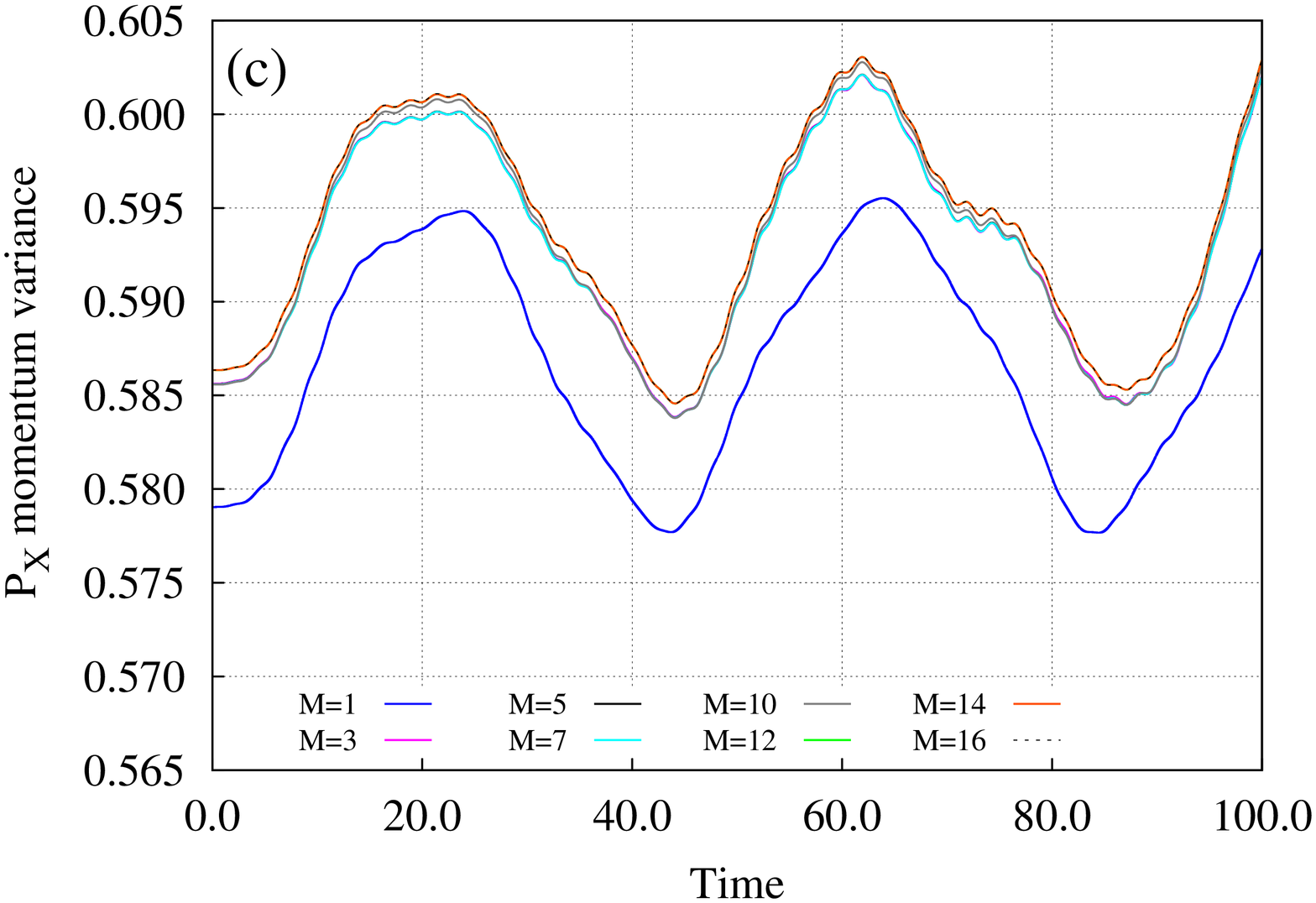}
\includegraphics[width=0.2930\columnwidth,angle=-90]{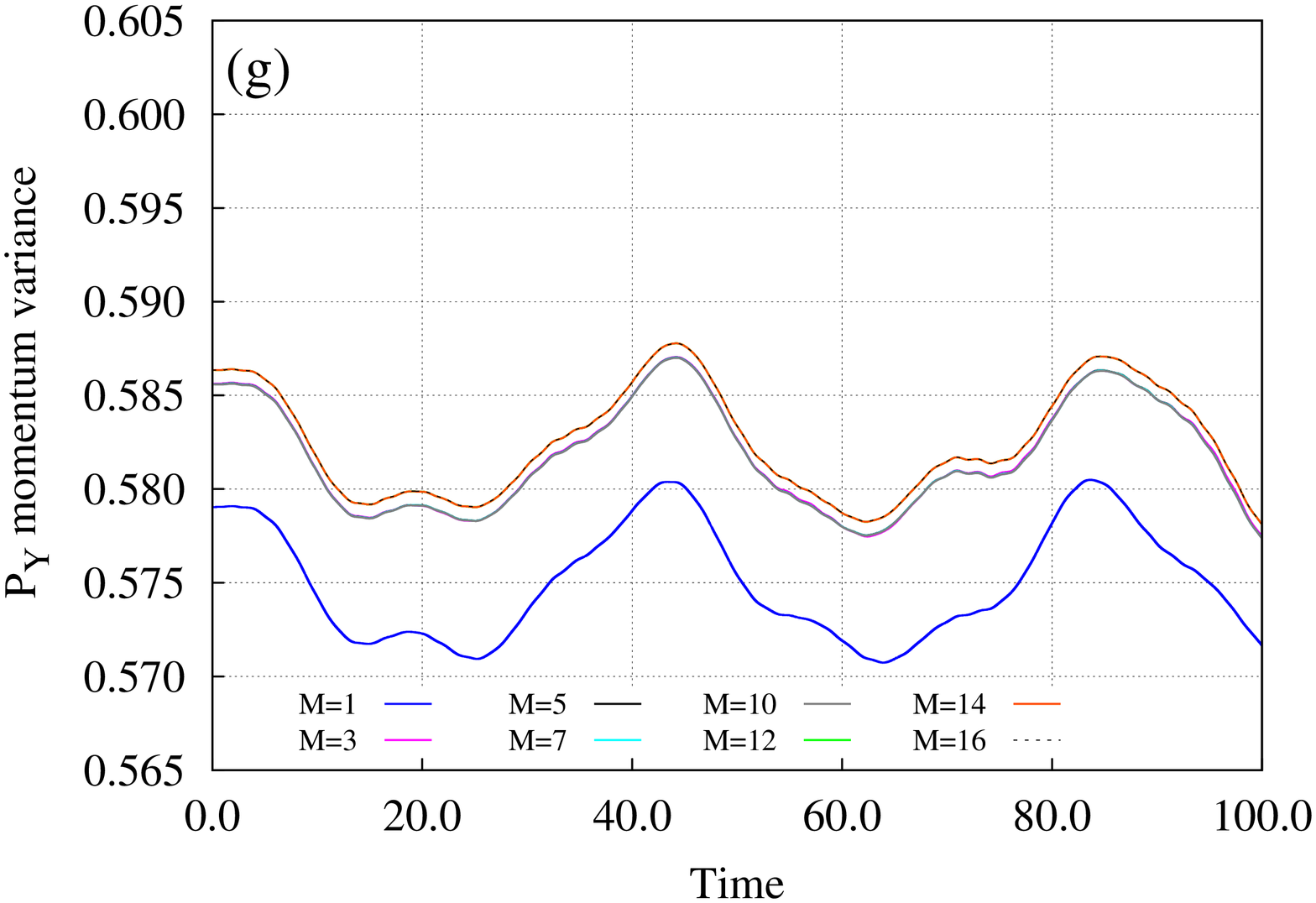}
\vglue 0.25 truecm
\hglue -1.0 truecm
\includegraphics[width=0.2930\columnwidth,angle=-90]{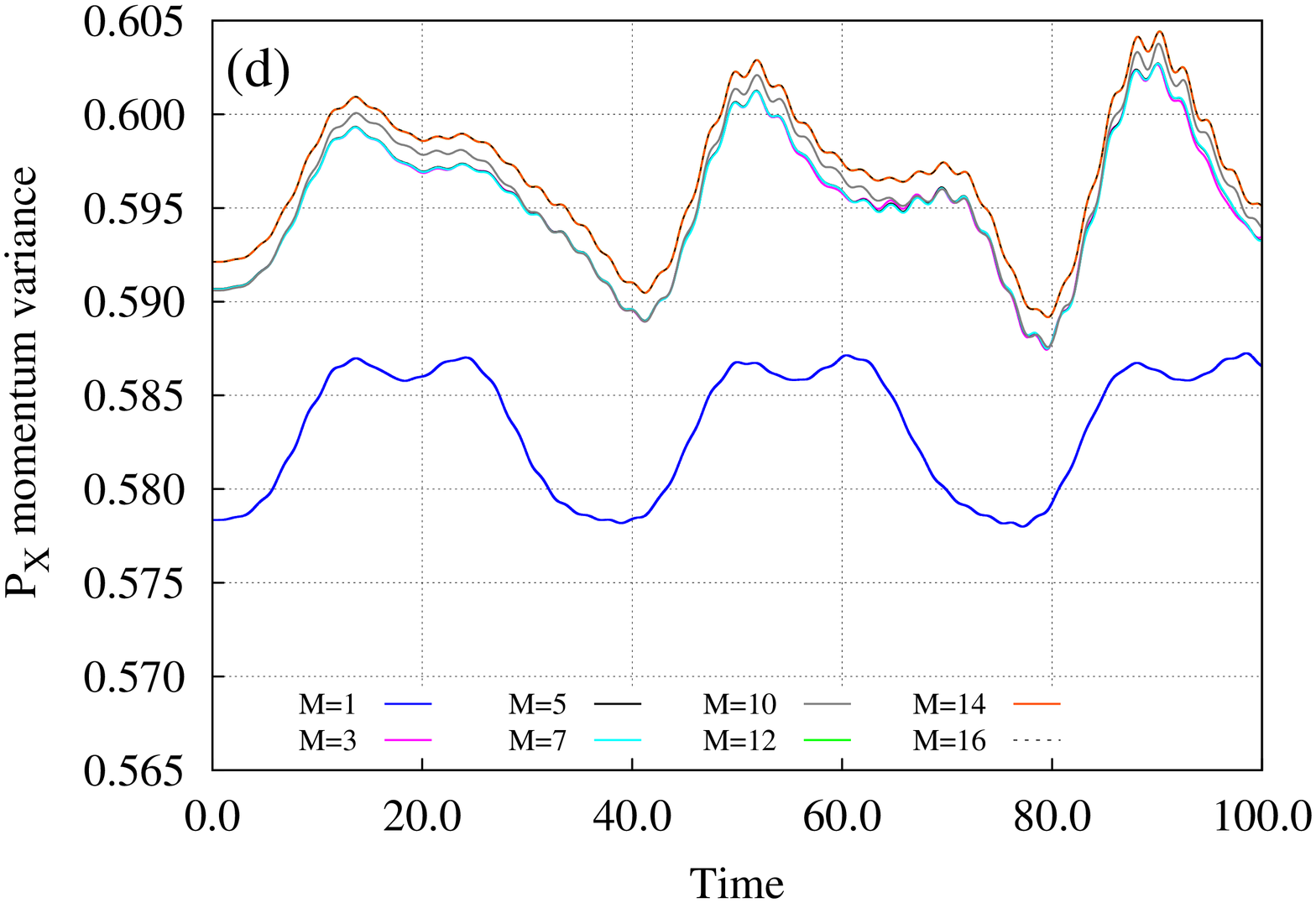}
\includegraphics[width=0.2930\columnwidth,angle=-90]{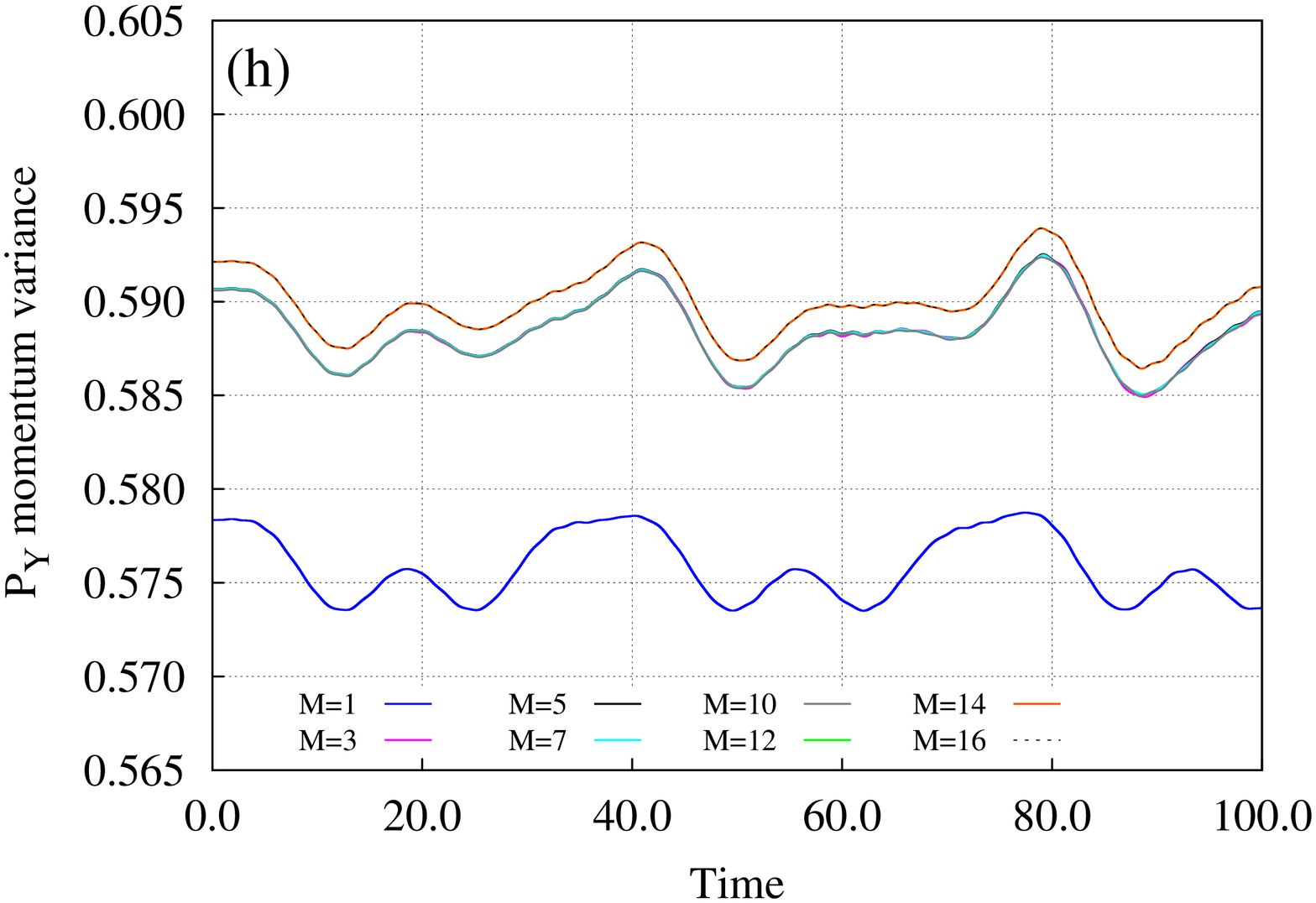}
\end{center}
\vglue 0.25 truecm
\caption{Momentum variance dynamics following a potential quench.
The mean-field ($M=1$ time-adaptive orbitals) and many-body (using $M=3$, $5$, $7$, $10$, $12$, $14$, and $15$, $16$ time-adaptive orbitals)
time-dependent momentum variances per particle,
$\frac{1}{N}\Delta^2_{\hat P_X}(t)$ [left column, panels (a), (b), (c), and (d)] and
$\frac{1}{N}\Delta^2_{\hat P_Y}(t)$ [right column, panels (e), (f), (g), and (h)],
of $N=10$ bosons in the annuli 
with barrier heights and interaction strengths
(a) and (d) $V_0=5$, $\lambda_0=0.02$;
(b) and (e) $V_0=5$, $\lambda_0=0.04$;
(c) and (f) $V_0=10$, $\lambda_0=0.02$; and 
(d) and (g) $V_0=10$, $\lambda_0=0.04$
following a sudden potential tilt by $0.01 x$.
The respective depletions are plotted in Fig.~\ref{f2}.
See the text for more details.
The quantities shown are dimensionless.}
\label{f4}
\end{figure}

We move to the momentum variance and also make contact with
the results of the position variance.
Fig.~\ref{f4} displays the many-particle momentum variance per particle, $\frac{1}{N} \Delta^2_{\hat P_X}(t)$ and $\frac{1}{N} \Delta^2_{\hat P_Y}(t)$,
for $V_0=5$, $V_0=10$ and $\lambda_0=0.02$, $\lambda_0=0.04$.
Just like the results of the position variance,
since rotational symmetry is lifted the dynamics of respective quantities along the x-axis and y-axis are different
[the initial conditions imply 
$\frac{1}{N} \Delta^2_{\hat P_X}=\frac{1}{N} \Delta^2_{\hat P_Y}$ at $t=0$].
The mean-field ($M=1$) and many-body ($M \ge 3$) values are, again, separated from each other,
but now the former lie below the later,
and there is only about $1\%$-$4\%$ of a difference 
depending on the repulsion strength and barrier height, also see \cite{INF7,var2}.
Thus, the momentum variance rather weakly depends on the (small amount of) depletion.
This is because the matrix elements in (\ref{VAR_GEN}) are
typically smaller with the momentum operator
than with the position operator. 
Yet, despite their small difference, 
the many-body and mean-field momentum variances do not cross each other, see Fig.~\ref{f4}
(contrast with the interaction-quench dynamics in a single trap in \cite{var1}).

It is instructive to analyze the momentum-variance dynamics at short times.
Whereas $\Delta^2_{\hat P_X}(t)$ primarily increases,
$\Delta^2_{\hat P_Y}(t)$ mainly decreases.
This matches the geometry of the sloshing dynamics in the tilted annulus,
in which bosons from the `north' and `south' poles (on the y-axis) 
start to move to
the left and accumulate in the `west' pole (on the x-axis),
and that the cross section of the rim of an annulus is enlarged when moving away from the center of the annulus. 
In other words,
the dynamics of the momentum variances at short times when moving to the left
reflects the relative localization of the bosons
in the $x$ direction and the effective broadening of the wavepacket 
along the $y$ direction.
Both the mean-field and many-body variances oscillate with a very small amplitude, note the scale on the y-axis in Fig.~\ref{f4}.
The high-frequency oscillations mark high-energy radial excitations across the (tight) annulus rim \cite{var2}.
Like for the position variance, 
the amplitude of oscillations of the y-axis momentum variances is smaller than that of the x-axis momentum variances.
Finally, we see that already $M=3$ orbitals accurately describe the dynamics of the momentum variance
(the difference to the $M>3$ results is lower than $1\%$).

\begin{figure}[!]
\begin{center}
\hglue -1.0 truecm
\includegraphics[width=0.345\columnwidth,angle=-90]{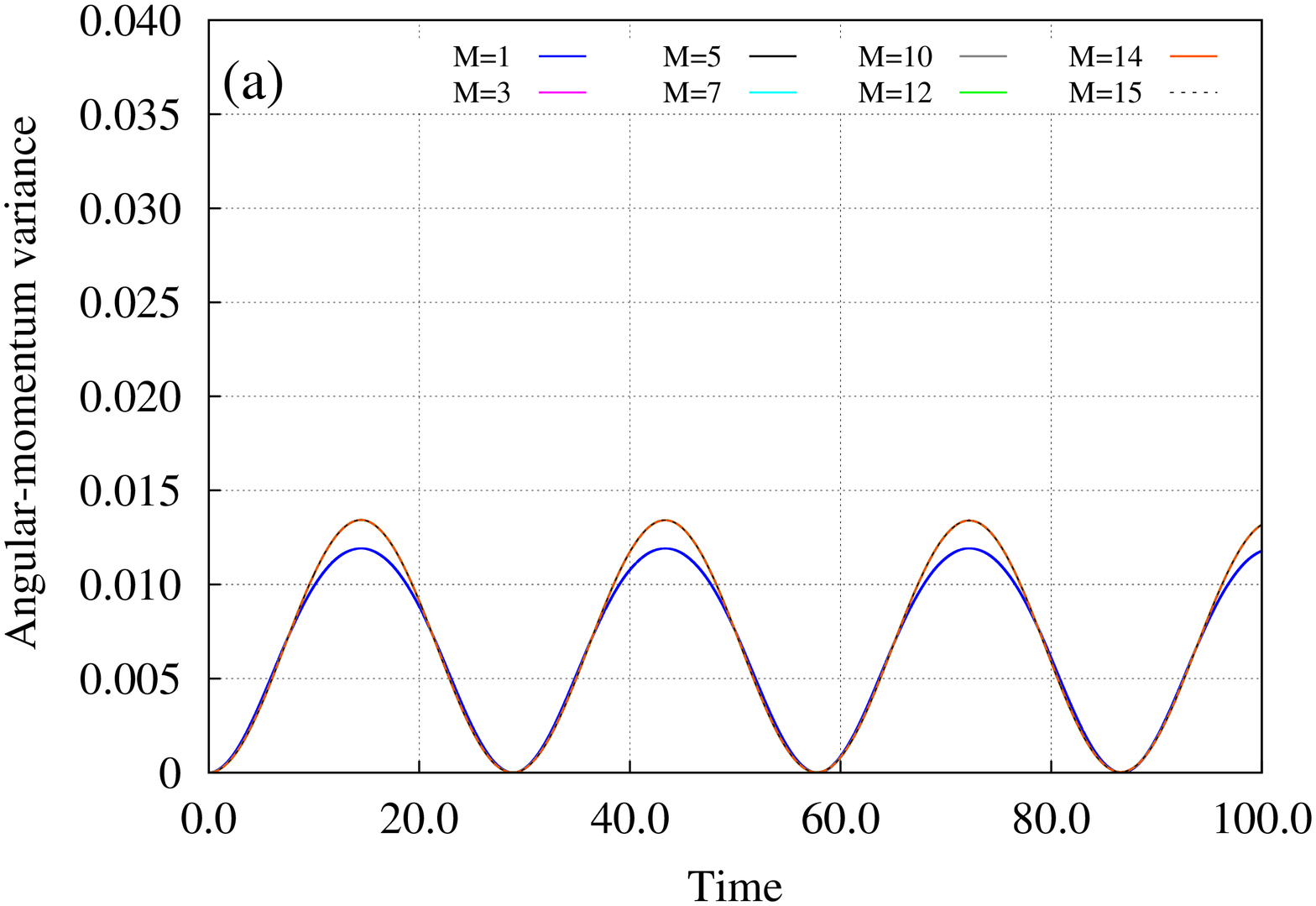}
\includegraphics[width=0.345\columnwidth,angle=-90]{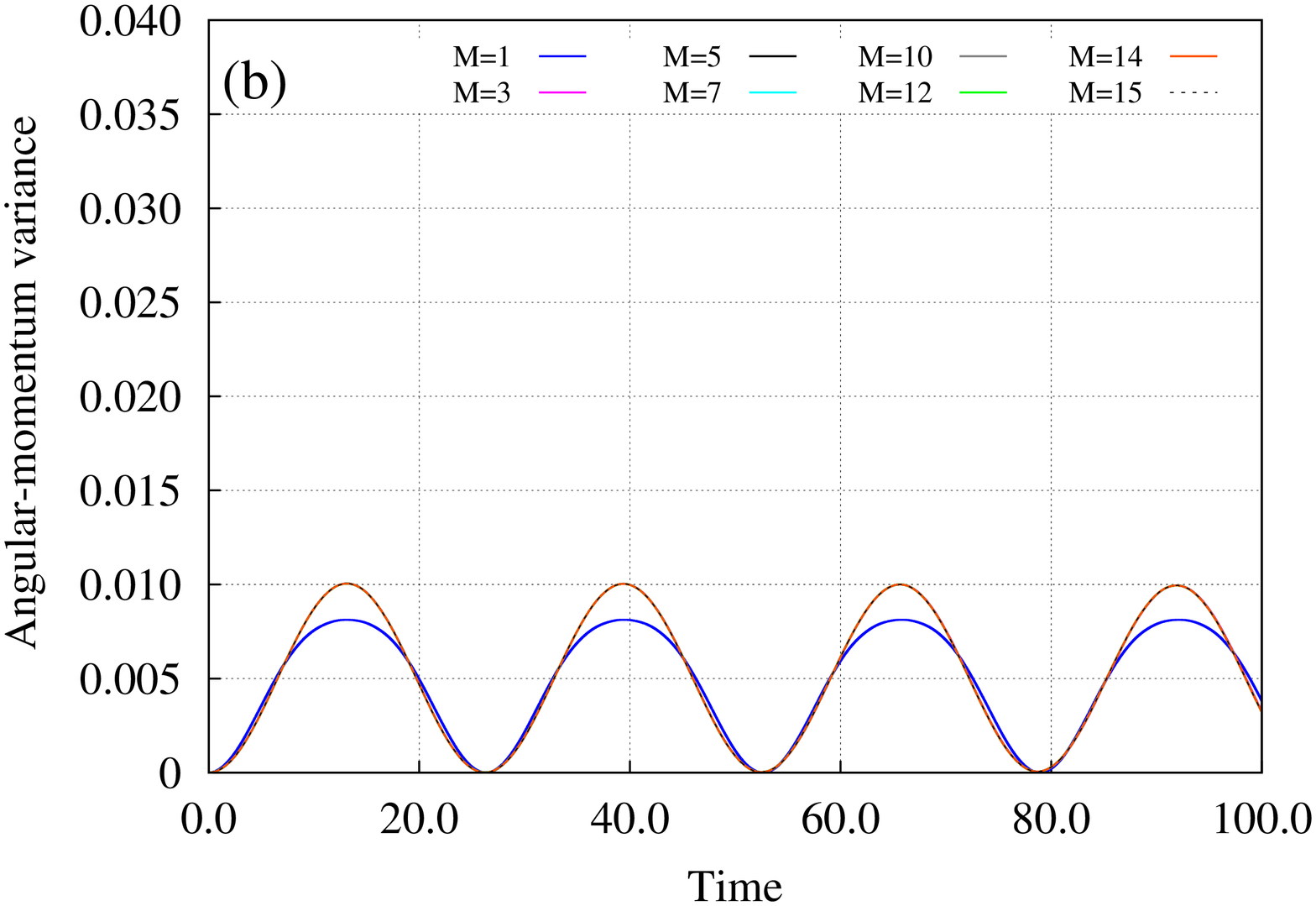}
\hglue -1.0 truecm
\includegraphics[width=0.345\columnwidth,angle=-90]{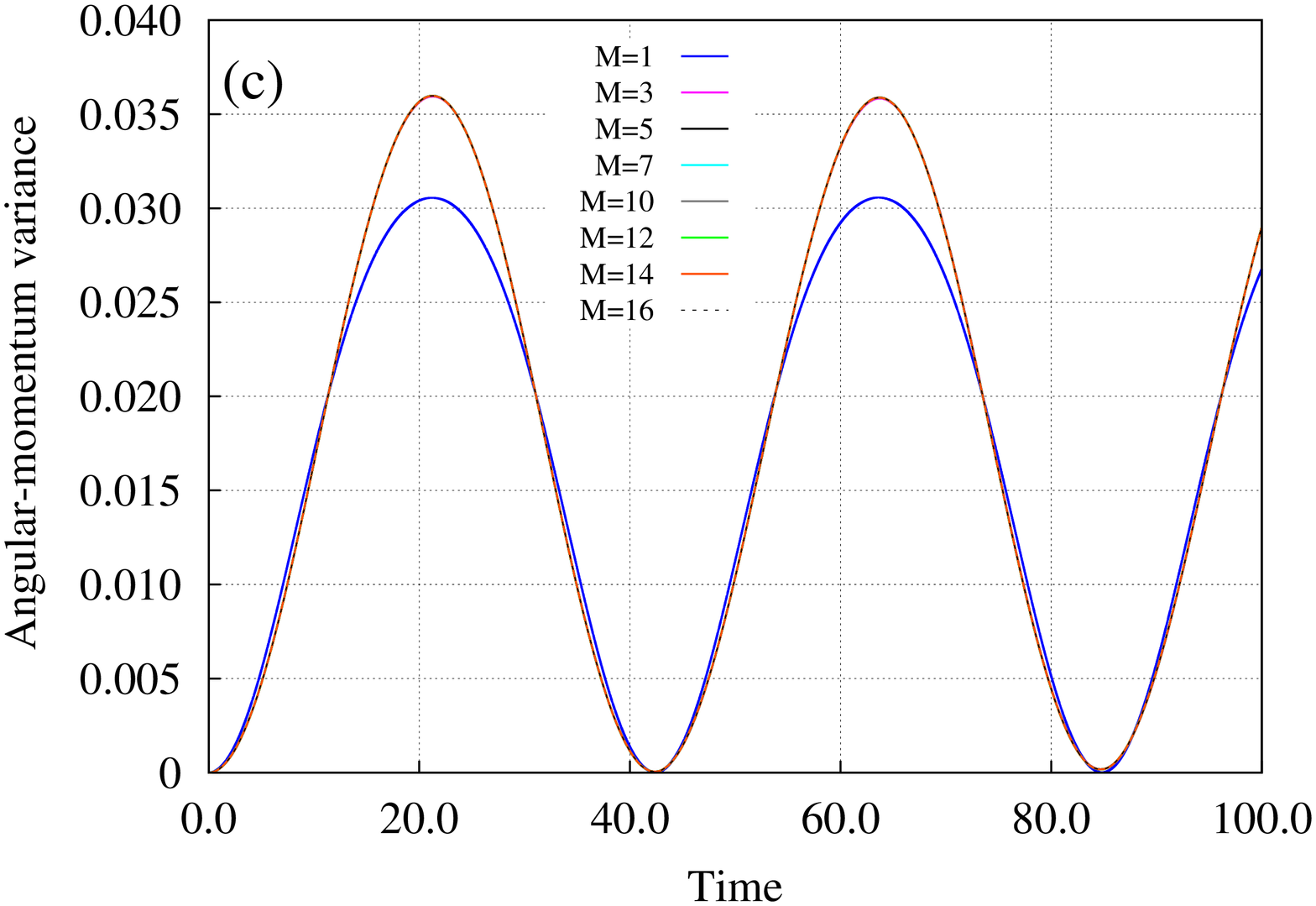}
\includegraphics[width=0.345\columnwidth,angle=-90]{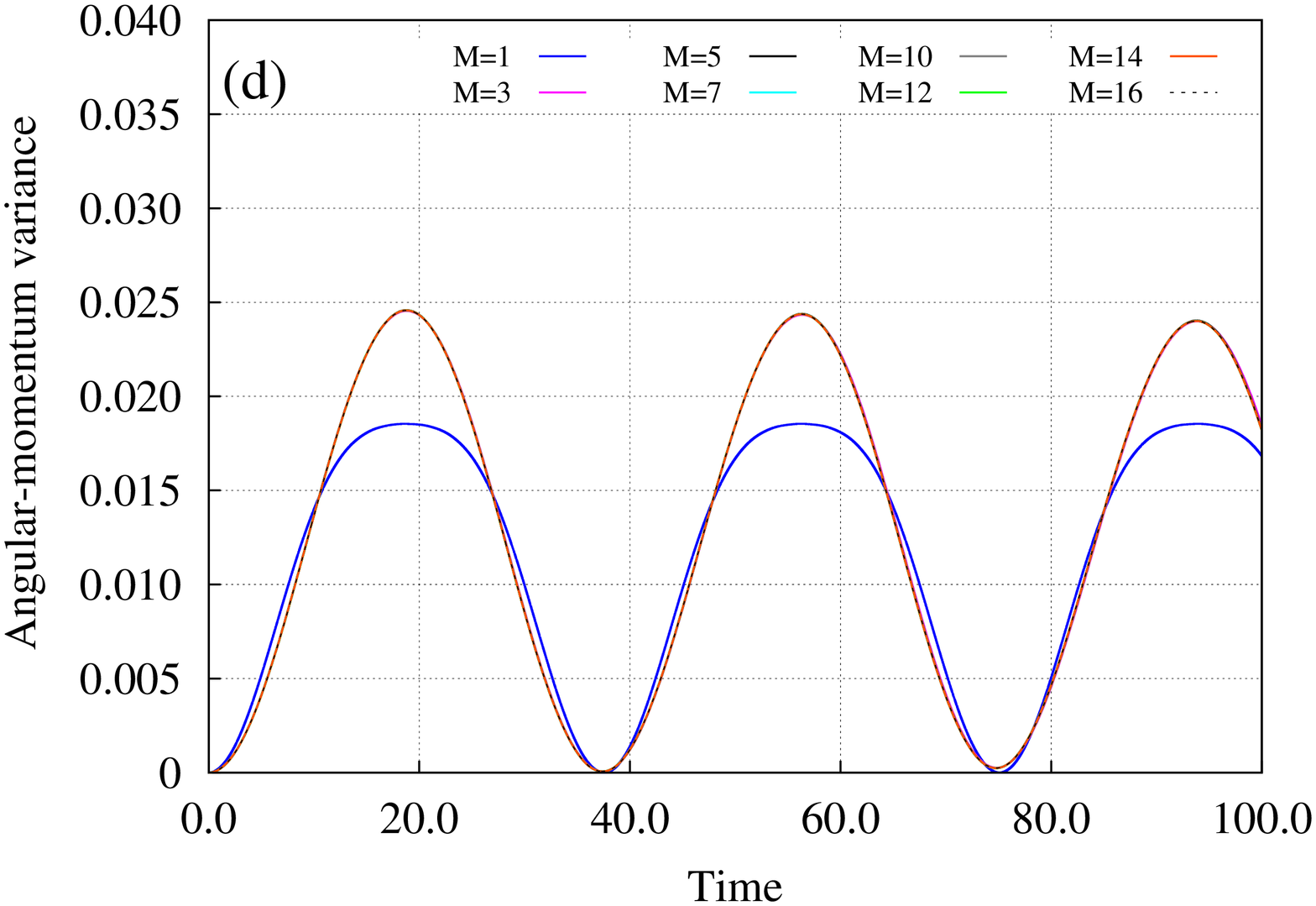}
\end{center}
\vglue 0.75 truecm
\caption{Angular-momentum variance dynamics following a potential quench.
The mean-field ($M=1$ time-adaptive orbitals) and many-body (using $M=3$, $5$, $7$, $10$, $12$, $14$, and $15$, $16$ time-adaptive orbitals)
time-dependent angular-momentum variance per particle,
$\frac{1}{N}\Delta^2_{\hat L_Z}(t)$,
of $N=10$ bosons in the annuli 
with barrier heights and interaction strengths
(a) $V_0=5$, $\lambda_0=0.02$;
(b) $V_0=5$, $\lambda_0=0.04$;
(c) $V_0=10$, $\lambda_0=0.02$; and 
(d) $V_0=10$, $\lambda_0=0.04$
following a sudden potential tilt by $0.01 x$.
The respective depletions are plotted in Fig.~\ref{f2}.
See the text for more details.
The quantities shown are dimensionless.}
\label{f5}
\end{figure}

We now move to the angular-momentum variance and an interesting inter-connection with the momentum variance.
Fig.~\ref{f5} presents the many-particle angular-momentum variance per particle, $\frac{1}{N} \Delta^2_{\hat L_Z}(t)$,
for the two barrier heights, $V_0=5$ and $V_0=10$,
and the two interaction strengths, $\lambda_0=0.02$ and $\lambda_0=0.04$.
There are several features seen in the dynamics.
Since rotational symmetry is lifted, $\frac{1}{N} \Delta^2_{\hat L_Z} \ne 0$
expect for the initial conditions at $t=0$
(the values of the minima for $t>0$, see below, are close to but not $0$).
The dynamics of $\frac{1}{N} \Delta^2_{\hat L_Z}(t)$ appears to be almost periodic and rather regular,
more than that for the respective position and momentum variances, compare to Figs.~\ref{f3} and \ref{f4}.
On the other end,
focusing on the dynamics of the center-of-mass in Fig.~\ref{f1},
one can clearly observe correlation between the two quantities;
Whenever $\frac{1}{N}\langle\Psi|\hat X|\Psi\rangle(t)$ has a minimum, i.e., the bosons are maximally localized to the left, 
$\frac{1}{N} \Delta^2_{\hat L_Z}(t)$ has a maximum,
and whenever $\frac{1}{N}\langle\Psi|\hat X|\Psi\rangle(t)$ has a maximum (which value is about $0$),
i.e., the bosons are momentarily, approximately equally distributed along the annulus,
$\frac{1}{N} \Delta^2_{\hat L_Z}(t)$ has a minimum (which value, as mentioned above, is close to $0$).
Furthermore, the frequencies of the two quantities as well as their relative amplitudes
as a function of the barrier height and interaction strength are alike.
These observations call for a dedicated analysis.

To shed light on the above dynamics of the angular-momentum variance, see Fig.~\ref{f5},
we analyze the translational properties of variances in Appendix \ref{VAR_TRANS}.
Whereas the position variances and, trivially, the momentum variances, are translationally invariant,
this invariance does not hold for the angular-momentum variance.
If a wavepacket prepared in the origin has angular-momentum variance $\frac{1}{N} \Delta^2_{\hat L_Z}$,
then several terms are added when the wavepacket is translated to the point $(a,b)$ in plane,
and angular-momentum variance is
thereafter computed, see Eq.~(\ref{Var_Lz_a_b}).
Now, if this wavepacket is rotationally symmetric, i.e., $\frac{1}{N} \Delta^2_{\hat L_Z}=0$,
then several of the terms in (\ref{Var_Lz_a_b}) vanish due to spatial symmetry and we are left 
with the appealing relation,  
$\frac{1}{N} \Delta^2_{\hat L_Z}\Big|_{\Psi(a,b)} = 
a^2 \frac{1}{N} \Delta^2_{\hat P_Y}\Big|_{\Psi} + 
b^2 \frac{1}{N} \Delta^2_{\hat P_X}\Big|_{\Psi}$ [Eq.~\ref{Var_Lz_Spher_Symm}],
connecting the angular-momentum variance of $\Psi(a,b)$ localized at $(a,b)$
and $\Psi$ at the origin.
The meaning of this relation is that the momentum variances, $\frac{1}{N} \Delta^2_{\hat P_X}$ and $\frac{1}{N} \Delta^2_{\hat P_Y}$,
together with the spatial translations along the y-axis and x-axis, respectively,
determine the angular-momentum variance of a translated wavepacket (rotationally-symmetric at the origin).

Returning to and combining Fig.~\ref{f5} for the angular-momentum variance, Fig.~\ref{f1} for the center-of-mass dynamics,
and Fig.~\ref{f4}e,f,g,h for $\frac{1}{N} \Delta^2_{\hat P_Y}$,
we can now discuss and explain their inter-connection.
Explicitly, the center-of-mass dynamics is analogous to translating the wavepacket along the x-axis (back and forth to the left),
hence, according to Eq.~(\ref{Var_Lz_Spher_Symm}), $\frac{1}{N} \Delta^2_{\hat P_Y}$ is needed.
This is why the dependencies of the frequency and amplitude of oscillations of $\frac{1}{N} \Delta^2_{\hat L_Z}(t)$
on the barrier height and interaction strength
nicely follow, respectively, those of $\frac{1}{N}\langle\Psi|\hat X|\Psi\rangle(t)$, compare Figs.~\ref{f5} and \ref{f1}.
What is the role of $\frac{1}{N} \Delta^2_{\hat P_Y}$ then?
The momentum variance helps us understand the deviations
between the many-body and mean-field results in Fig.~\ref{f5}.
We see that the maxima of the many-body $\frac{1}{N} \Delta^2_{\hat L_Z}(t)$ ($M>3$)
are larger than the maxima of the mean-field $\frac{1}{N} \Delta^2_{\hat L_Z}(t)$ ($M=1$).
The difference is about $7\%$-$25\%$ (compare to the low depletion, Fig.~\ref{f2}),
depending of $V_0$ and $\lambda_0$,
and follows the respective trend of the many-body and mean-field results for $\frac{1}{N} \Delta^2_{\hat P_Y}$,
see Fig.~\ref{f4}e,f,g,h.
We note that, although the wavepacket describing the bosons dynamics in the tilted annulus
is not a translated, rotationally-invariant wavepacket,
and the values of deviations (in percents) between the
many-body and mean-field results are actually larger for 
$\Delta^2_{\hat L_Z}$ (at the maxima) than for $\Delta^2_{\hat P_Y}$,
we find the above analytically-based analysis to well explain
the numerical findings and trends.
Last but not least, 
a close inspection of the many-body and mean-field curves of the angular-momentum
variance in Fig.~\ref{f5} shows that there are instances when they cross each other,
i.e., one is smaller or larger than the other.
This is in contrast with the non-crossing of the many-body and mean-field position and momentum variances,
see Figs.~\ref{f3} and \ref{f4}, respectively.
Finally,
we find that already $M=3$ time-adaptive orbitals
accurately describe the dynamics of the angular-momentum variance.

\begin{figure}[!]
\begin{center}
\hglue -1.0 truecm
\includegraphics[width=0.456\columnwidth,angle=-90]{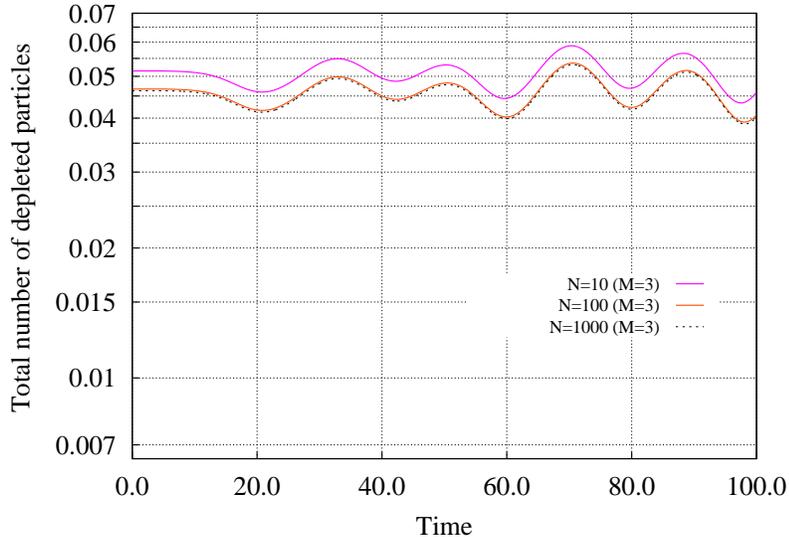}
\end{center}
\vglue 0.75 truecm
\caption{Depletion dynamics following a potential quench {\it en route} to the infinite-particle limit.
The time-dependent total number of depleted particles, $N-n_1(t)$,
of $N=10$, $N=100$, and $N=1000$ bosons with interaction parameter $\Lambda=\lambda_0(N-1)=0.36$
for an annulus with barrier height $V_0=10$ following a sudden potential tilt by $0.01 x$.
The number of time-adaptive orbitals is $M=3$.
The respective position, momentum, and angular-momentum
variances along with the expectation value  of the center-of-mass 
are plotted in Fig.~\ref{f7}.
See the text for more details.
The quantities shown are dimensionless.
}
\label{f6}
\end{figure}

Our investigations are nearing their end,
what is left to explore is the behavior of the position, momentum, and angular-momentum variances 
at the infinite-particle limit.
Which of the above-described detailed findings, plotted in Figs.~\ref{f1}-\ref{f5}
for a rather small ($N=10$ bosons) yet weakly-depleted BEC,
survive this limit?
To answer the question,
we concentrate on the system with the higher barrier, $V_0=10$,
and stronger interaction (for $N=10$ bosons), $\lambda_0=0.04$.
We hence fix the interaction parameter $\Lambda=\lambda_0(N-1)=0.36$,
and compute and compare
the dynamics for $N=10$, $N=100$, and $N=1000$ bosons
using $M=3$ time-adaptive orbitals.
We have seen for $N=10$ bosons that $M=3$ time-adaptive orbitals accurately describe the variances.
This implies that, keeping the interaction parameter $\Lambda$ fixed  
while increasing the number of particles $N$,
using $M=3$ time-adaptive orbitals for calculating the variances will be (at least) 
as accurate as for $N=10$ particles,
see in this respect \cite{conv_2}.
Before we proceed, a methodological remark.
Examining the convergence of properties with the number of particles
for $N=10$, $N=100$, and $N=1000$
bosons is (still) far away from infinity, see in this respect \cite{INF10}.
We hence use, interchangeably,
the term {\it en route} to the infinite-particle limit.
We shall see below that, in effect, the infinite-particle
limit is practically well achieved for the variances already 
for $N=1000$ bosons. 

\begin{figure}[!]
\begin{center}
\vglue -1.25 truecm
\hglue -1.0 truecm
\includegraphics[width=0.345\columnwidth,angle=-90]{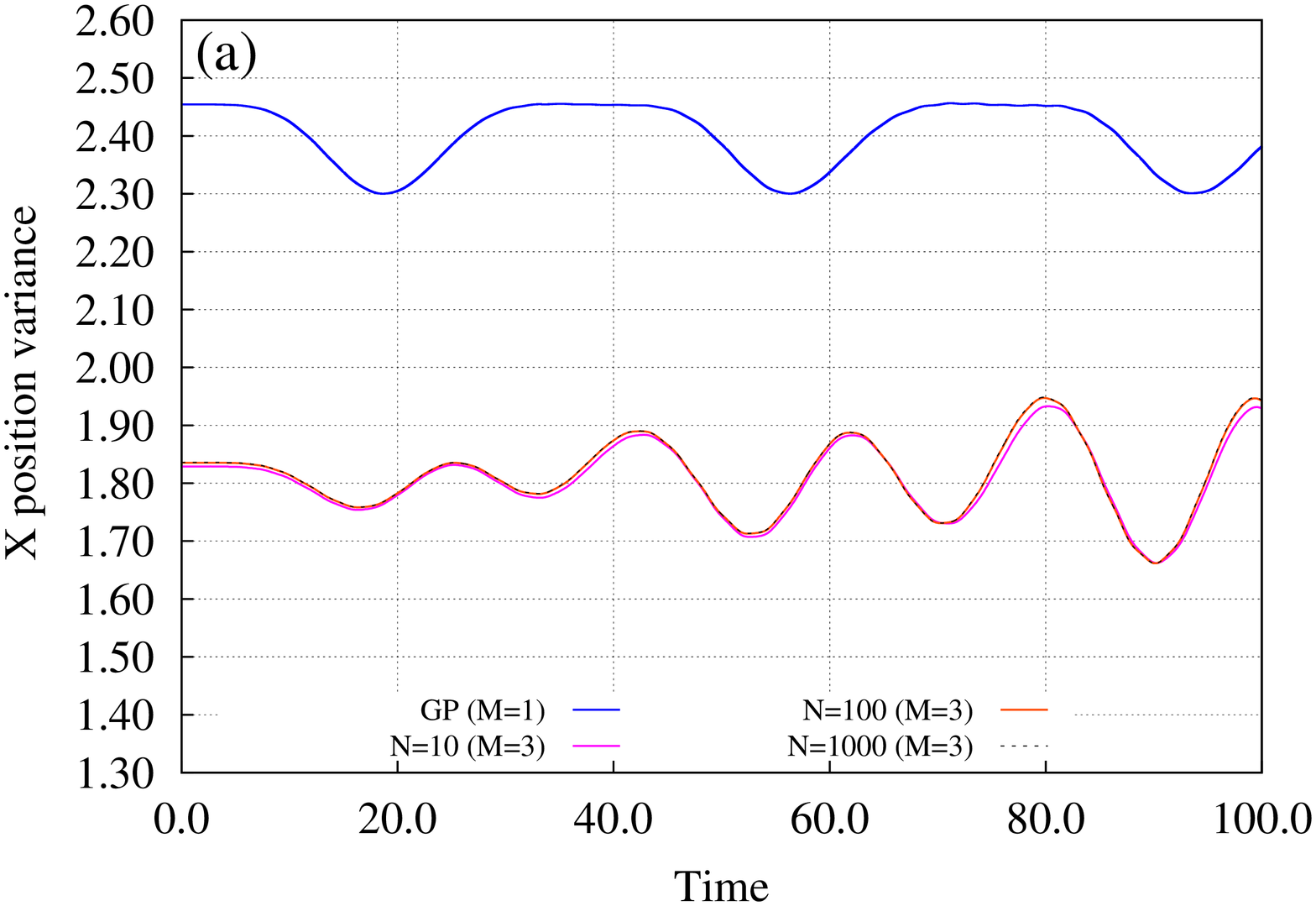}
\includegraphics[width=0.345\columnwidth,angle=-90]{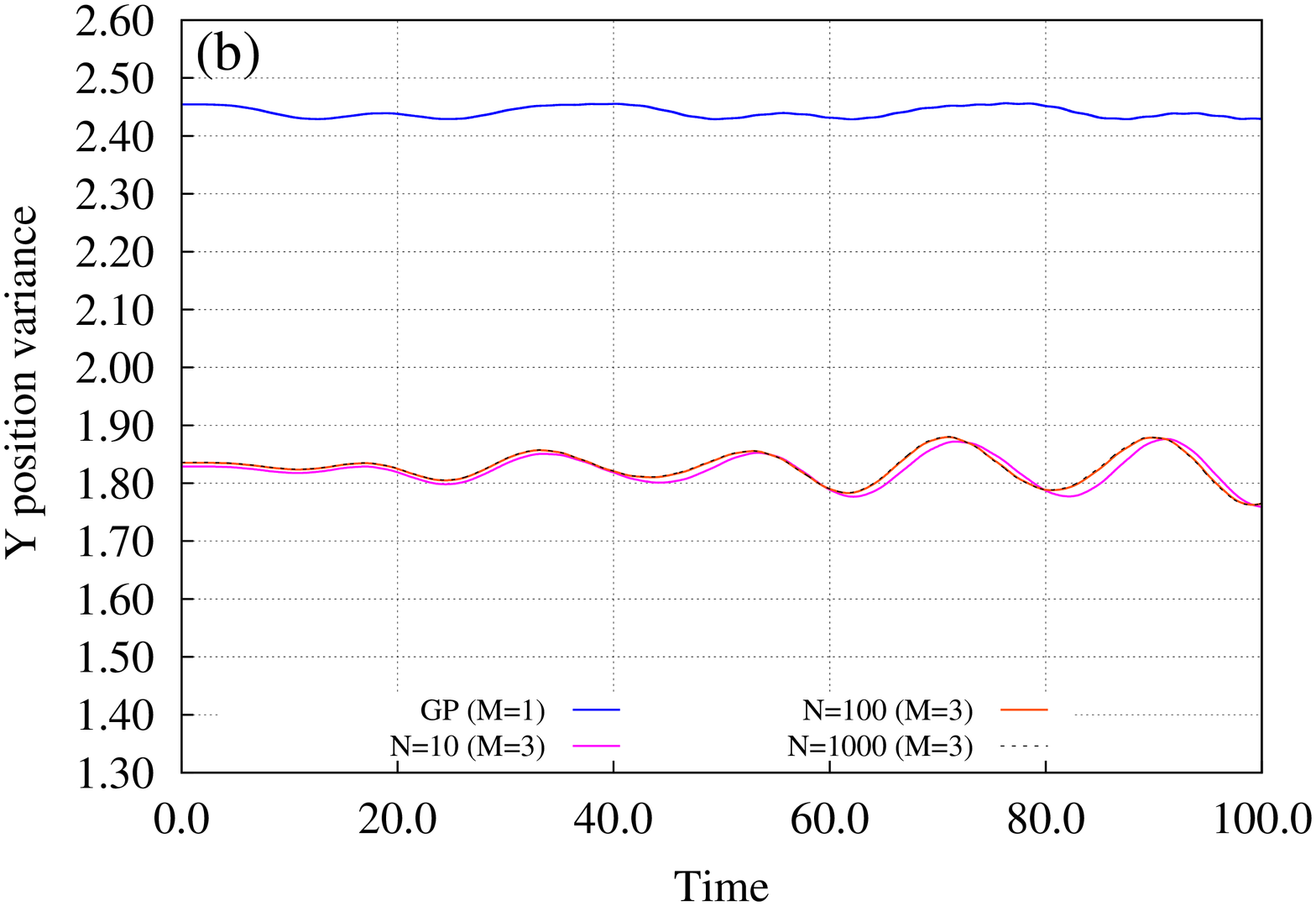}
\vglue 0.25 truecm
\hglue -1.0 truecm
\includegraphics[width=0.345\columnwidth,angle=-90]{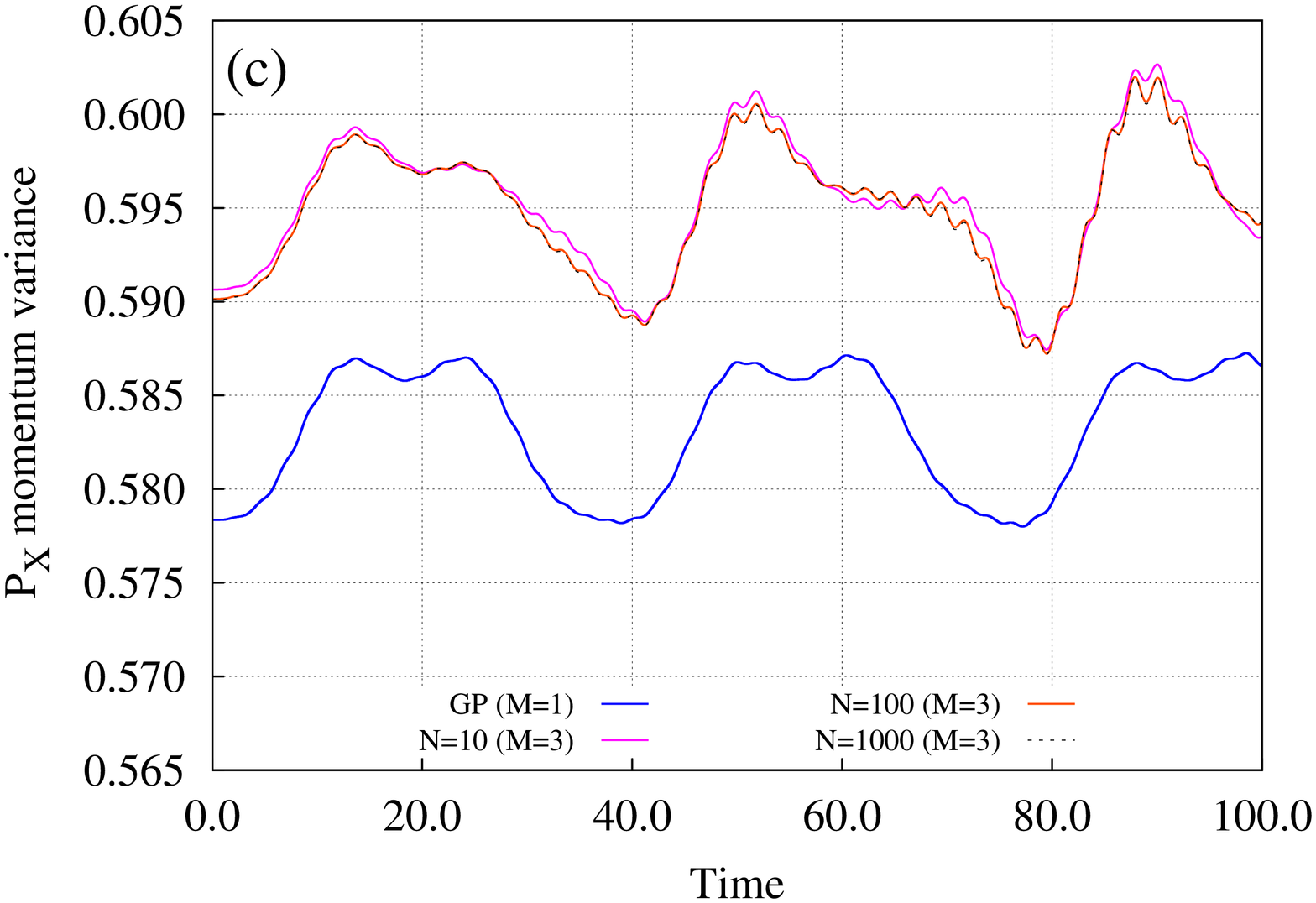}
\includegraphics[width=0.345\columnwidth,angle=-90]{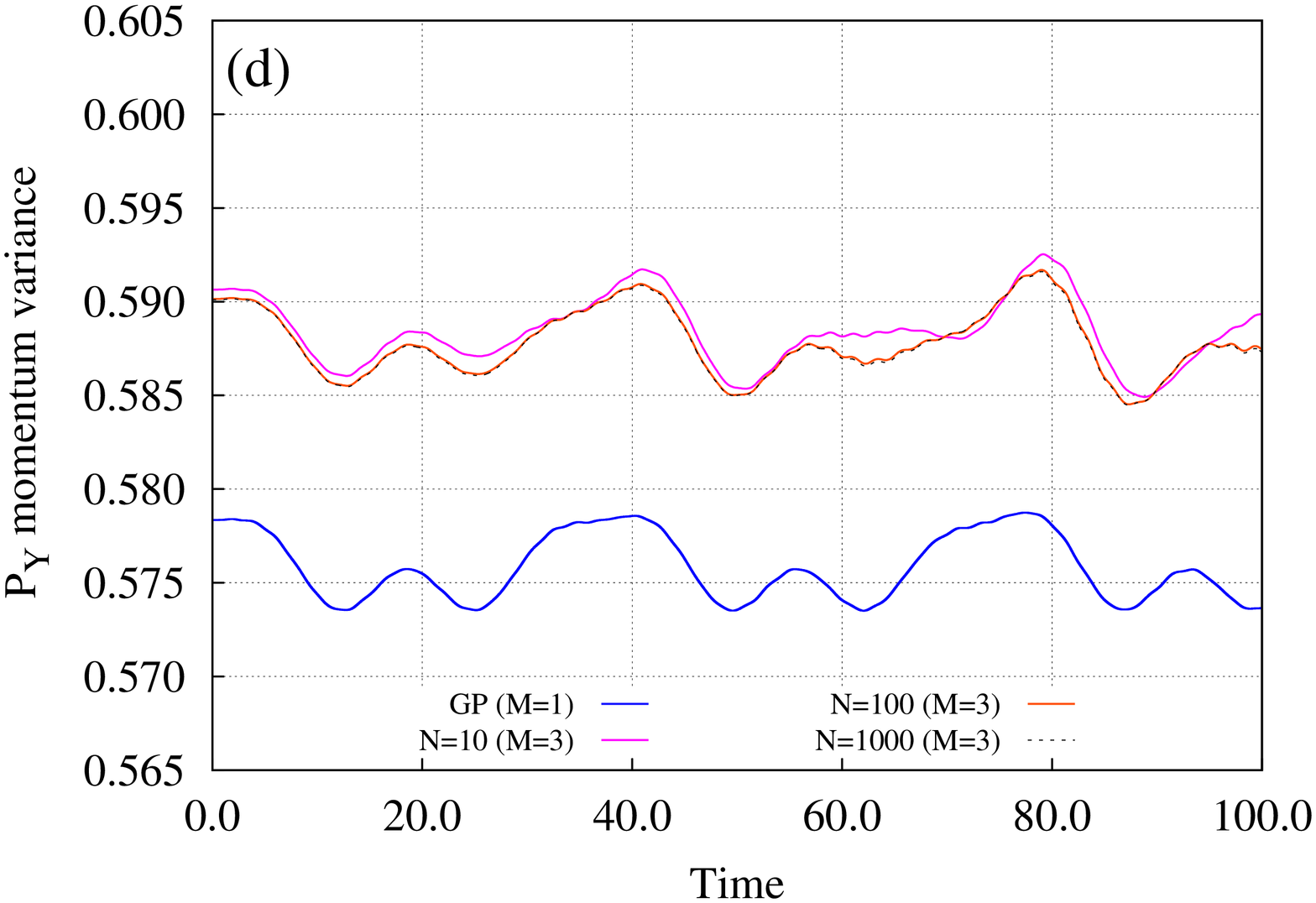}
\vglue 0.25 truecm
\hglue -1.0 truecm
\includegraphics[width=0.345\columnwidth,angle=-90]{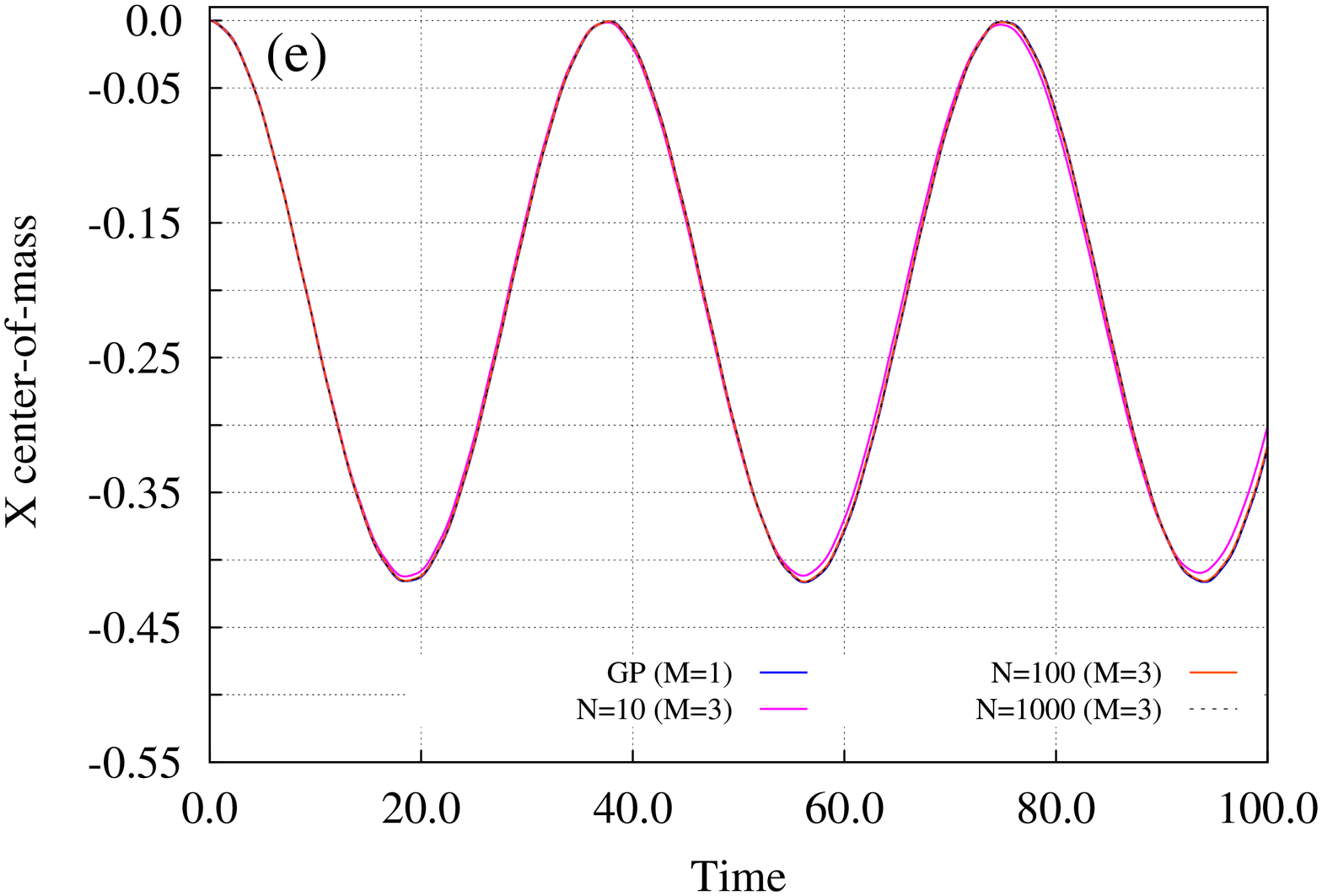}
\includegraphics[width=0.345\columnwidth,angle=-90]{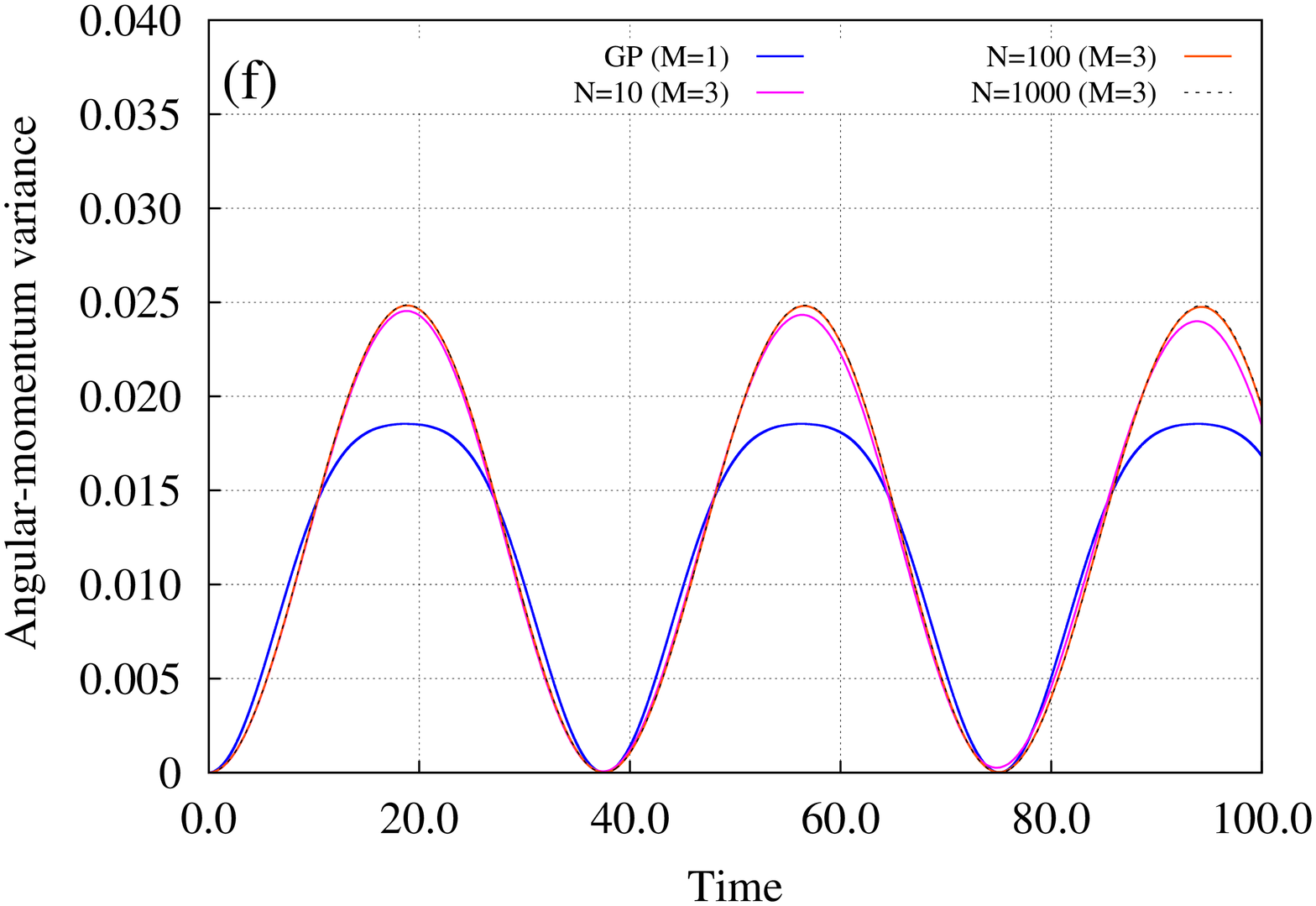}
\end{center}
\vglue 0.25 truecm
\caption{Position, momentum, and angular-momentum variance dynamics following a potential quench {\it en route} to the infinite-particle limit.
The mean-field ($M=1$ time-adaptive orbitals) and many-body (using $M=3$ time-adaptive orbitals)
time-dependent position variances per particle, (a) $\frac{1}{N}\Delta^2_{\hat X}(t)$ and (b) $\frac{1}{N}\Delta^2_{\hat Y}(t)$,
momentum variances per particle, (c) $\frac{1}{N}\Delta^2_{\hat P_X}(t)$ and (d) $\frac{1}{N}\Delta^2_{\hat P_Y}(t)$,
and angular-momentum variance per particle, (f) $\frac{1}{N}\Delta^2_{\hat L_Z}(t)$,
of $N=10$, $N=100$, and $N=1000$ bosons with interaction parameter $\Lambda=\lambda_0(N-1)=0.36$
for an annulus with barrier height $V_0=10$ following a sudden potential tilt by $0.01 x$.
(e) The time-dependent expectation value of the center-of-mass, $\frac{1}{N}\langle\Psi|\hat X|\Psi\rangle(t)$. 
The respective depletions are plotted in Fig.~\ref{f6}.
See the text for more details.
The quantities shown are dimensionless.}
\label{f7}
\end{figure}

Fig.~\ref{f6} prints the total number of depleted particles, $N-n_1(t)$, for $N=10$, $N=100$, and $N=1000$
bosons for $\Lambda=0.36$ and $V_0=10$ using $M=3$ time-adaptive orbitals.
Convergence of the number of depleted particles with $N$ is nicely seen.
Since $N-n_1(t)$ converges to a finite (and small) value with $N$,
the bosons are becoming $100\%$ condensed in the limit of an infinite number of particles,
i.e., $\frac{n_1(t)}{N} \to 1$ as $N \to \infty$ (at least up to the maximal time of the computation, $t=100$).

Fig.~\ref{f7} exhibits the position variances per particle, 
$\frac{1}{N}\Delta^2_{\hat X}(t)$ and $\frac{1}{N}\Delta^2_{\hat Y}(t)$,
momentum variances per particle,
$\frac{1}{N}\Delta^2_{\hat P_X}(t)$ and $\frac{1}{N}\Delta^2_{\hat P_Y}(t)$,
angular-momentum variance per particle, $\frac{1}{N}\Delta^2_{\hat L_Z}(t)$,
and the expectation value of the
center-of-mass, $\frac{1}{N}\langle\Psi|\hat X|\Psi\rangle(t)$, 
for $N=10$, $N=100$, and $N=1000$
bosons and
for $\Lambda=0.36$ and $V_0=10$ using $M=3$ time-adaptive orbitals.
Once again, convergence of each of the quantities with $N$ is clearly seen.
Yet, whereas the center-of-mass dynamics
converges to the mean-field dynamics when the number of particles is increased,
the variances exhibit many-body dynamics which converges nicely with $N$,
but not to the respective mean-field dynamics.
Beyond that,
all the above results,
for the frequencies, amplitudes, anisotropies, inter-connections,
and particularly the differences between the many-body and mean-field
position, momentum, and angular-momentum variances
persist at the limit of infinite number of particles,
despite the bosons becoming
$100\%$ condensed.
This brings the present analysis to an end.

\section{Summary and outlook}\label{SUMM}

In the present work 
we studied, analytically and numerically, 
the position, momentum, and especially the angular-momentum variance 
of interacting bosons trapped in a two-dimensional anisotropic trap
for static and dynamic scenarios.
The differences between the variances at the mean-field level,
which are attributed to the shape of the density per particle,
and the respective variances at the many-body level,
which incorporate a small amount of depletion outside the condensed mode, 
were used to characterize sometimes large
position, momentum, and angular-momentum
correlations in the BEC
for finite systems and at the limit of an infinite number of particles
where the bosons are $100\%$ condensed.
Finally, we also explored and utilized inter-connections between
the variances,
particularly between the angular-momentum and momentum variances, 
through the analysis of their translational properties.

There are many intriguing directions to follow out of which we list three below.
First, variances of BECs in the rotating frame of reference
in which high-lying excitations become low-energy excitations and even the ground state.
Second, angular-momentum variance of a BEC flowing past an obstacle
in which the mean angular-momentum variance vanishes.
And third, variances in three-dimensional geometries lacking lower-dimensional analogs,
such as a M\"obius strip.
In all these cases, whether considering a few interacting bosons or a BEC in the infinite-particle limit,
interesting exciting results are expected.

\section*{Acknowledgements}

This research was supported by the Israel Science Foundation (Grant No. 600/15).
We thank Kaspar Sakmann,
Anal Bhowmik, Sudip Haldar, and Raphael Beinke for discussions.
Computation time on the BwForCluster,
the High Performance Computing system Hive of the Faculty of Natural Sciences at University of Haifa,
and the Cray XC40 system Hazelhen at the High Performance Computing Center
Stuttgart (HLRS) is gratefully acknowledged.

\appendix
\section{Variances and translations}\label{VAR_TRANS}

Consider the many-particle translation operator in two spatial dimensions
$e^{-i(\hat P_X a + \hat P_Y b)}$, where $\hat P_X=\sum_{j=1}^N \hat p_{x,j}$ and $\hat P_Y=\sum_{j=1}^N \hat p_{y,j}$.
Its operation on a multi-particle wavefunction $\Psi$ is given by $e^{-i(\hat P_X a + \hat P_Y b)} \Psi(x_1,y_1,\ldots,x_N,y_N)=\Psi(x_1-a,y_1-b,\ldots,x_N-a,y_N-b) \equiv \Psi(a,b)$.
What are the implications on the variances when computed with respect to the translated wavefunction $\Psi(a,b)$?

For the position operator
$\hat X=\sum_{j=1}^N \hat x_j$ (and equivalently for $\hat Y=\sum_{j=1}^N \hat y_j$)
we have $\langle\Psi(a,b)| \hat X |\Psi(a,b)\rangle = \langle\Psi| \hat X |\Psi\rangle +N a$
and $\langle\Psi(a,b)| \hat X^2 |\Psi(a,b)\rangle = \langle\Psi| \hat X^2 |\Psi\rangle + 2 N a \langle\Psi| \hat X |\Psi\rangle + (N a)^2$,
implying that
\beq
\frac{1}{N} \Delta^2_{\hat X}\Big|_{\Psi(a,b)} = \frac{1}{N} \Delta^2_{\hat X}\Big|_{\Psi}.
\eeq
Trivially for the momentum operator
$\hat P_X$ (and equivalently for $\hat P_Y=\sum_{j=1}^N \hat p_{y,j}$) one has
\beq
\frac{1}{N} \Delta^2_{\hat P_X}\Big|_{\Psi(a,b)} = \frac{1}{N} \Delta^2_{\hat P_X}\Big|_{\Psi},
\eeq
i.e., both the position variance and momentum variance are translationally invariant.

For the angular-momentum operator
$\hat L_Z=\sum_{j=1}^N \left(\hat x_j \hat p_{y,j} - \hat y_j \hat p_{x,j}\right)$ the situation is more interesting.
From $\langle\Psi(a,b)| \hat L_Z |\Psi(a,b)\rangle = 
\langle\Psi| \hat L_Z |\Psi\rangle + a \langle\Psi| \hat P_Y |\Psi\rangle - b \langle\Psi| \hat P_X |\Psi\rangle$\break\hfill and
$\langle\Psi(a)| \hat L_Z^2 |\Psi(a)\rangle = 
\langle\Psi| \hat L_Z^2 |\Psi\rangle + a^2 \langle\Psi| \hat P_Y^2 |\Psi\rangle + b^2 \langle\Psi| \hat P_X^2 |\Psi\rangle +
a \langle\Psi| \hat L_Z \hat P_Y + \hat P_Y \hat L_Z |\Psi\rangle -
b \langle\Psi| \hat L_Z \hat P_X + \hat P_X \hat L_Z |\Psi\rangle -
2 a b \langle\Psi| \hat P_Y \hat P_X|\Psi\rangle$ 
we have
\beqn\label{Var_Lz_a_b}
& & \frac{1}{N} \Delta^2_{\hat L_Z}\Big|_{\Psi(a,b)} = 
\frac{1}{N} \Delta^2_{\hat L_Z}\Big|_{\Psi} + 
a^2 \frac{1}{N} \Delta^2_{\hat P_Y}\Big|_{\Psi} + 
b^2 \frac{1}{N} \Delta^2_{\hat P_X}\Big|_{\Psi} + 
\nonumber \\
& & \qquad + a \left( \langle\Psi| \hat L_Z \hat P_Y + \hat P_Y \hat L_Z |\Psi\rangle - 
2 \langle\Psi| \hat L_Z |\Psi\rangle \langle\Psi| \hat P_Y |\Psi\rangle \right) - \nonumber \\
& & \qquad - b \left( \langle\Psi| \hat L_Z \hat P_X + \hat P_X \hat L_Z |\Psi\rangle - 
2 \langle\Psi| \hat L_Z |\Psi\rangle \langle\Psi| \hat P_X |\Psi\rangle \right) - \nonumber \\
& & \qquad - 2ab \left( \langle\Psi| \hat P_Y \hat P_X |\Psi\rangle - \langle\Psi| \hat P_Y |\Psi\rangle \langle\Psi| \hat P_X |\Psi\rangle \right). \
\eeqn
Eq.~(\ref{Var_Lz_a_b}) deserves a discussion.
In turn,
even for the ground state of an interacting many-boson system in a rotationally-symmetric
[for which $\frac{1}{N}\Delta^2_{\hat L_Z}=0$ holds]
but otherwise translated trap,
the angular-momentum variance
\beq\label{Var_Lz_Spher_Symm}
\frac{1}{N} \Delta^2_{\hat L_Z}\Big|_{\Psi(a,b)} = 
a^2 \frac{1}{N} \Delta^2_{\hat P_Y}\Big|_{\Psi} + 
b^2 \frac{1}{N} \Delta^2_{\hat P_X}\Big|_{\Psi} 
\eeq
differs at the many-body level and mean-field level of theory,
i.e.,
when $a, b \ne 0$ and $\lambda_0 \ne 0$. 
This is, as can be seen in (\ref{Var_Lz_Spher_Symm}),
because of the respective many-body and mean-field
momentum variances, $\frac{1}{N}\Delta^2_{\hat P_X}$ and $\frac{1}{N}\Delta^2_{\hat P_Y}$.
The analytical result (\ref{Var_Lz_Spher_Symm})
is employed to analyze the numerical findings for the
time-dependent angular-momentum variance in the main text.
Generally in the absence of spatial symmetries, see Eq.~(\ref{Var_Lz_a_b}),
more terms contribute to the translated angular-momentum variance.

\end{document}